# Self Modulation and Scattering Instability of a Relativistic Short Laser Pulse in an Underdense Plasma

J. Yazdanpanah

*Plasma and Nuclear Fusion Research School, Nuclear Science and Technology Research Institute, Tehran, Iran*

**Abstract**

Characterization of self-consistent laser-plasma evolutions serves as a fundamental issue in the field of relativistic laser-plasma interactions. In this paper, we present an analysis framework for description of these evolutions during propagation of a short intense laser pulse in a sub-critical high-density plasma (the pulse length exceeds the plasma wavelength). In this context, the pulse evolutions are attributed to the wakefield induced self-modulation and destabilization via parametric exponentiation of the initial noise content. The self-consistent plasma evolutions are formulated in terms of quantities which used to be motion constants in the absence of pulse evolutions. This proves very useful both in understanding plasma evolutions during self-modulation and also in facilitating the instability studies in the strongly nonlinear regime, via refinement of unstable plasma perturbations. General analytical solutions, at arbitrary pulse conditions, are derived for self-modulation, indicating that the envelop evolutions are driven by the induced spatial frequency-chirp. Also, these results state that the envelope attains fine modulations which produce long wavelength low-frequency modes via beating the carrier mode. The plasma wave variations are found to convect and amplify away from the pulse front. Regarding parametric instability, we assess different scattering regimes at different pulse shapes and peak intensities, manifesting anomalous behaviors ranging from wild positioning of the Stokes wave in dispersion plane to broadening in the scattered spectrum and halting the instability. Our analyses are assisted and verified by numerous fluid and particle-in-cell simulations. Based on our results, we discuss phenomena like the pulse breakup and its different regimes, and assisted particle acceleration in presence of pulse evolutions.

jamyazdan@gmail.com, jyazdanpanah@aeoi.org.ir

I. Introduction

Nowadays a unique opportunity is provided by modern high-intensity short pulse lasers, for experimental investigation of many fundamental phenomena [1], and realization of very attractive applied concepts like laser-plasma accelerator (LPA) [2] and fast-ignition scheme of inertial confinement fusion (ICF) [3]. These investigations and applications most often imply or contain the propagation of a short intense laser pulse through a high-density transparent plasma [1-17]. This is the case for example in the self-modulated laser wakefield accelerator (SMLWFA) [9,10], and in laser interactions with solid targets [11,12] and ICF targets [3,4,11], which usually leads to production of coronal plasmas by pulse pedestal before arriving the main part of the pulse. We may also add, laser interaction with modern foam targets, which have attracted many recent attentions [14, 15].

The short pulse propagation inside the high-density plasma is characterized by the regime for which the pulse length, $L_p$, exceeds the initial plasma wavelength, $c/\omega_{pe}$ where $\omega_p = \sqrt{n_{e0}e^2/\varepsilon_0 m_e}$ is the plasma frequency and $c$ is the light speed (and $n_{e0}$, $\varepsilon_0$, $e$, $m_e$ are respectively the plasma density, the vacuum permittivity, elementary charge and electron mass), viz. the condition $L_p > c/\omega_{pe}$ implies $n_{e0} > \varepsilon_0 m_e c^2 / e^2 L_p^2$. This regime has significant differences with the so-called ultra-short pulse regime (the area of the laser wakefield accelerator (LWFA) [2]) regarding the stability of the laser pulse propagation inside the plasma. Despite the latter, in the former the laser pulse is susceptible to scattering instabilities [16-20] and pulse breakup [17, 21-24]. In most applications, these effects on the laser pulse act as a double-edged sword; they both restrict the pulse penetration into the plasma [25, 26], and in the same time lead to enhanced particle acceleration [27-34]. Generally, scattering assisted electron acceleration is now considered as a generic mechanism for production of very-energetic/super-ponderomotive electron populations observed in high density plasma irradiation by intense pulses [13, 27, 30-34]. In SMLWFA, the electron acceleration is also enhanced via the wakefield amplification by the pulse breakup [27]. In the context of ICF, it is most often intended to control the instabilities and pulse breakup to enhance the laser penetration into the target [4], but in the same time, production of energetic electrons may be considered as a mechanism for attaining the electron ignition-beam [13]. Therefore, careful characterization of such pulse evolutions is crucial for their control and possibly their optimized utilization.

The pulse breakup phenomenon has been extensively studied in previous works [16,17, 21-24], and has been subjected to many debates regarding its origins [17]. *Antonsen and Mora* [16], and Sprangle etal [21] have described this phenomenon as an adiabatic sausaging process originated from the transverse plasma oscillations. Mori *etal* [17], on the other hand, have attributed this phenomenon to the direct forward Raman scattering (FRS), a quite longitudinal mechanism. Following *Mori etal*, Gordon *etal* [35] (see also *Mima etal* [36]), have proposed a two stage process; in the first phase, shortly after the laser entrance, the laser experiences Raman backward scattering (RBS) instability which seeds the plasma wave and the subsequent Raman forward scattering instability. In the second phase, the pulse breakup continues via seeded RFS, as described by *Mori etal* [17].

In past two decades, many efforts have also been made on parametric instabilities of intense and/or short laser pulses [16-20, 37-42]. It should be emphasized that in these studies the wake excitation has been usually ruled out by assuming infinite pulse lengths or low intensities. In this regard, we may mention the works by *Barr etal* [41, 42] which apparently considered the most relevant conditions to the present study, say, relativistic short pulse interaction. Even in these works, the wake excitation has been implicitly ignored via assumption of a homogenous pump wave in Lorentz boosted frame. For a real short pulse, though experiencing Lorentz elongation (inverse Fitzgerald-contraction), the pulse length remains of finite length in the boosted farm, and the wakefield may not be ignored. A treatment of short laser pulses in the boosted frame is presented by *Yazdanpanah* [43] which reveals the importance of induced wakefied on pulse evolutions.

In this paper, we present an analysis framework for description of self-consistent laser-plasma evolutions during a short intense laser pulse propagation in a high density plasma (pulse length exceeds the plasma wave length). The most distinguishing feature of the intense short pulse interactions, compared to the common cases of infinite pulse length and subrelativistic intensities, is the possible presence of initial strong wakefield. This in turns leads to faster development of self-modulation with respect to FRS. In this regard, the pulse evolutions are attributed to the wakefield induced self-modulation accompanied by destabilization via parametric exponentiation of the initial noise content (scattering instabilities). Especially, we describe the *longitudinal* pulse breakup in terms of self-modulation rather than FRS considered by *Mori etal* [17] and *Gordon etal* [35]. In addition, we recover the interplay between the initial wake excitation and parametric instabilities, which remained unresolved in previous studies [16-20, 37-42].

We formulate the plasma wave in terms of quantities which used to be motion constants in the absence of pulse evolutions, say electron energy and electron flux in the Galilean commoving window (not Lorentz boosted frame). This not only reveals the important dynamics of these quantities in the presence of pulse evolutions, but also, as will be seen, facilitates instability studies in the presence of strong wakefield via refinement of unstable field-plasma perturbations.

In the case of self-modulation, the approach under-consideration is an extension of works done by *Schroeder etal* [44] and *Yazdanpanah* [43] on *ultra-short* pulses (pulse lengths shorter than plasma wavelength), to include pulse lengths exceeding the plasma wavelength. Furthermore, by applying a technical mathematical treatment, the obtained self-modulation equations are solved and reduce to closed results which well agree with presented simulation results. In the case of instabilities, after assessment of different regimes, we employ numerous fluid and particle-in-cell (PIC) simulations to accomplish and verify our analytical arguments.

The organization of our paper is as follow; basic equations are summarized in Sec. (II). In Sec. III, we describe pulse self-modulation and self-consistent plasma motion. In Sec. (IV), we describe parametric instabilities. In Sec. V, we summarize the numerical simulations and discuss our results. Finally, in Sec. VI, we point out our conclusions and remarks.

## II. Basic equations

A vast number of nonlinear phenomena in the intense laser interaction with the under-dense plasma are investigated by applying the well-known set of cold fluid plus Maxwell equations (see e.g. [2] and references therein). Here we consider these equations to examine the interaction of a plan p-polarized laser pulse with a uniform plasma slab (one dimensional geometry), where the laser propagates along $x$ axis and is polarized in $y$ direction. Therefore, after transforming into Pulse Co-Moving Window (PCMW) ($(x,t) \rightarrow (\xi \equiv x - v_g t, t)$), our basic equations in terms of normalized quantities read as:

$$\frac{\partial n_e}{\partial t} + c \frac{\partial}{\partial \xi}[n_e(v_{ex} - v_g)] = 0 \tag{1a}$$

$$(\frac{\partial}{\partial t} - cv_g \frac{\partial}{\partial \xi}) p_{ex} = -\frac{\partial}{\partial \xi}(\gamma_e - \phi) \tag{1b}$$

$$p_{ey} = A_y \tag{1c}$$

$$\frac{\partial^2 \phi}{\partial \xi^2} = \frac{\omega_p^2}{c^2}[n_e - 1] \tag{1d}$$

$$\left[\frac{1}{\gamma_g^2}\frac{\partial^2}{\partial \xi^2} - \frac{1}{c^2}\frac{\partial^2}{\partial t^2} + 2\frac{v_g}{c}\frac{\partial^2}{\partial \xi \partial t}\right]A_y = \frac{\Omega_p^2}{c^2}A_y \qquad (1e)$$

Here, $n_e$, $\mathbf{p}_e$, $\mathbf{v}_e = \mathbf{p}_e/\gamma_e$, $\gamma_e = (1-\mathbf{v}_e\cdot\mathbf{v}_e)^{-1/2}$, $\phi$, $\mathbf{A}$, $c$, $v_g = ck_0/\omega_0$ and $\gamma_g = (1-v_g^2)^{-1/2}$ stand for electron density, electron momentum, electron velocity, electron gamma factor, scalar potential, vector potentials, light speed and linear group velocity and its relativistic gamma factor, respectively; $k_0$ and $\omega_0$ are carrier wave-number and frequency. We have also defined $\Omega_p^2 \equiv \omega_p^2 n_e/\gamma_e$ where $\omega_p = (n_{e0}e^2/\varepsilon_0 m_e)^{1/2}$ is the plasma frequency, and $n_{e0}$, $\varepsilon_0$, $e$, $m_e$ are respectively the initial electron density, the vacuum permittivity, elementary charge and electron mass. All plasma and field quantities have been normalized in above equations but *length and time remained unnormalized*; we have set $e\mathbf{A}/m_e c \to \mathbf{A}$, $e\phi/m_e c^2 \to \phi$, $n_e/n_{e0} \to n_e$, $\mathbf{p}_e/m_e c \to \mathbf{p}_e$, $\mathbf{v}_e/c \to \mathbf{v}_e$ and $v_g/c \to v_g$.

It is well known that in the absence of temporal pulse evolutions in the PCMW, the following motion-constants are readily obtained from (1a) and (1b), respectively (see e.g. [45]),

$$J_x \equiv n_e(v_{ex} - v_g) = -v_g, \qquad (2a)$$

$$H_x \equiv \gamma_e - v_g p_{ex} - \phi = 1. \qquad (2b)$$

Generally, the laser pulse undergoes temporal variations in PCMW, and $J_x$ and $H_x$ do not remain further as motion constants. In the so-called quasi-static approximation (QSA) they are supposed to evolve very slowly in time, according to slow evolutions of the laser pulse (see e.g. [2, 45]). Quite generally, irrespective of $J_x$ and $H_x$ being constant or not, every plasma quantity may be expressed in terms these variables and vector and scalar potentials, i.e. we may make a transformation from the set of variables $\{n_e, p_{xe}, \phi, A_y\}$ to $\{J_x, H_x, \phi, A_y\}$. The procedure of this transformation is just already given in the well-known literatures for constant $J_x$ and $H_x$, i.e. for $J_x = -v_g$ and $H_x = 1$ [45]. For $v_g \simeq 1$ the final results become very simplified as summarized e.g. in [45]. Here, we consider $v_g \neq 1$ in the limit of $\gamma_g \gg 1$, viz. $v_g = 1 - 1/2\gamma_g^2$. Applying this approximation in (2a) and (2b) these equations respectively read as,

$$J_x^* \equiv n_e(v_{ex} - 1) \tag{2c}$$

$$H_x^* \equiv \gamma_e - p_{ex} - \phi \tag{2d}$$

where $J_x^* \equiv J_x - n_e^{(0)}/\gamma_g^2$ and $H_x^* \equiv H_x - p_{ex}^{(0)}/\gamma_g^2$, in which (0) suffix indicates the solutions when $v_g = 1$ be assumed (the leading order solution). In this way, for our frequently used quantities, we would have,

$$v_{ex} = \frac{\gamma_\perp^2 - (\phi + H_x^*)^2}{\gamma_\perp^2 + (\phi + H_x^*)^2} \tag{3a}$$

$$\gamma_e = \frac{1}{2}\left(\frac{\gamma_\perp^2}{\phi + H_x^*} + \phi + H_x^*\right) \tag{3b}$$

$$n_e = -J_x^* \frac{\gamma_\perp^2 + (\phi + H_x^*)^2}{2(\phi + H_x^*)^2} \tag{3c}$$

where $\gamma_\perp \equiv \sqrt{1 + A_y^2}$.

In incorporation of pulse evolutions, we will find it very useful to rewrite Eqs. (1a) and (1d) in terms of $J_x$ and $H_x$ which, now, are not motion constants. To do this, we firstly notice that Eq. (1b) may be easily brought into the following form,

$$\frac{\partial \gamma_e}{\partial t} = -cv_{ex}\frac{\partial}{\partial \xi}[\gamma_e - \phi - v_g p_{ex}] + \frac{A_y}{\gamma_e}\frac{\partial A_y}{\partial t}, \tag{4}$$

and this result may be combined with Eq. (1b) to give,

$$\frac{dH_x}{dt} = -\frac{\partial \phi}{\partial t} + \frac{A_y}{\gamma_e}\frac{\partial A_y}{\partial t} \tag{5}$$

where the complete time derivative has its usual meaning; $d/dt \equiv \partial/\partial t + cv_{ex}\partial/\partial x = \partial/\partial t + c(v_{ex} - v_g)\partial/\partial \xi$. On the other hand, we may combine the continuity equation (1a) with equations (1b) and (4), and respectively obtain,

$$\frac{\partial}{\partial t}[n_e p_{ex}] + c\frac{\partial}{\partial \xi}[n_e(v_{ex}-v_g)p_{ex}] = -cn_e\frac{\partial \phi}{\partial \xi} + c\frac{n_e A_y}{\gamma_e}\frac{\partial A_y}{\partial \xi}, \qquad (6a)$$

$$\frac{\partial}{\partial t}[n_e\gamma_e] + c\frac{\partial}{\partial \xi}[n_e(v_{ex}-v_g)\gamma_e] = n_e c v_{ex}\frac{\partial \phi}{\partial \xi} + \frac{n_e A_y}{\gamma_e}(\frac{\partial A_y}{\partial t} - cv_g\frac{\partial A_y}{\partial \xi}). \qquad (6b)$$

Now if we subtract (6b) multiplied by $v_g$ from (6a), we get, after some straightforward manipulations,

$$\gamma_e \frac{dJ_x}{dt} + J_x\frac{d\gamma_e}{dt} + c\gamma_e J_x\frac{\partial}{\partial \xi}[v_{ex}-v_g] = -n_e\{A_y(\frac{c}{\gamma_g^2}\frac{\partial A_y}{\partial \xi} + v_g\frac{\partial A_y}{\partial t}) + c(v_g v_{ex}-1)\frac{\partial \phi}{\partial \xi}\} \qquad (7a)$$

Eq. (7a) may be dramatically simplified and converted into a more useful form by expanding $d\gamma_e/dt$ as $d\gamma_e/dt = \partial \gamma_e/\partial t + c(v_{ex}-v_g)\partial \gamma_e/\partial \xi$ and rearranging its different terms. After some mathematical manipulations, using the identities $v_g^2 = 1-1/\gamma_g^2$ and $A_y\partial A_y/\gamma_e\partial \xi = \partial \gamma/\partial \xi - v_{ex}\partial p_{ex}/\partial \xi$, and finally substituting $\partial \gamma_e/\partial t$ from (4), we obtain,

$$\frac{dJ_x}{dt} + c\frac{n_e}{\gamma_e\gamma_{ex}^2}\frac{\partial H_x}{\partial \xi} = -\frac{n_e v_{ex}}{\gamma_e^2}A_y\frac{\partial A_y}{\partial t} \qquad (7b)$$

where $\gamma_{ex}$ is defined as $\gamma_{ex} = (1-v_{ex}^2)^{-1/2}$. Now, Eqs. (5) and (7b) together with Eqs. (1d) and (1e) (assisted by auxiliary relations (3a)-(3c)) form a complete set of equations for investigation of self-consistent laser-plasma evolutions in the fully nonlinear regime. Eqs. (5) and (7b) determine $H_x$ and $J_x$ dynamics which turn very important in particle acceleration in SMLWFA. In addition, they reveal an interesting property when subjected to instability studies, as will be seen in Sec. IV; linearization of these equations around the quasi static solutions leads to equations very similar to coupled oscillator equations inferred in the quasi-linear regime. In other words a refinement in field-plasma perturbations is attained using these equations.

Using the obtained equation we examine both the laser pulse self-modulation via wake excitation, and its destabilization via exponentiation of the initial noise content. In Sec. III, the self modulation is treated as a slow deformation in the pulse envelope and the set of equations (5), (7b), (1d) and (1e) (assisted by auxiliary relations (3a)-(3c)) are solved within the slow envelope approximation (SEA). In Sec. IV these equations are used to investigate propagation destabilization. This process is considered as a fast phenomena produced by coupling between the strong, low-frequency (in PCMW) laser field as the pump wave, and a high frequency electromagnetic mode from noise content, similar to the usual Raman instability.

### III. The pulse self modulation and self-consistent plasma motion

#### a. Pulse evolutions

We consider the wakefiled induced pulse self-modulations within the slow envelope approximation (SEA) [2, 21, 22, 43, 44, 46, 47]. In this regard, it should be mentioned that SEA has been frequently applied to study ultra-short [21, 43, 44, 46, 47] and short (longer than plasma wavelength) [21, 22] pulse interactions with plasma. Especially, we should mention the more recent work by *Schroeder etal* [44] on nonlinear evolutions of ultra-short laser pulses in plasma, which our treatment in this section is a generalization of this work to include general pulse lengths (short and ultra-short). We will find that laser frequency attains spatial chirp immediately after pulse entrance into the plasma, and that amplitude evolutions are driven dominantly by this frequency chirp.

By substitution of $A_y$ in terms of its envelope function, being denoted by $\hat{A}$, the wave equation (1e) may be rewritten as,

$$\left[\frac{1}{\gamma_g^2}\frac{\partial^2}{\partial \xi^2} - \frac{1}{c^2}\frac{\partial^2}{\partial t^2} + 2\frac{v_g}{c}\frac{\partial^2}{\partial \xi \partial t} + \frac{2i\omega_0}{c^2}\frac{\partial}{\partial t}\right]\hat{A} = \frac{\Omega_p^2 - \omega_p^2}{c^2}\hat{A} \qquad (8)$$

where $\hat{A}$ satisfies $A_y(\xi,t) = (\hat{A}(\xi,t)/2)e^{i(k_0 x - \omega_0 t)} + c.c.$, and $\omega_0$ and $k_0$ are respectively carrier frequency and wave-number. It is worth mentioning that by applying the linear dispersion relation, $\omega^2(k) = c^2 k^2 + \omega_p^2$, into the phase factor in the envelope presentation we get $k_0 x - \omega_0 t = k_0 \xi - (\omega_0 - c k_0 v_g)t \equiv k_0 \xi - \omega_{w0} t$. This relation defines the carrier frequency, $\omega_{w0}$, in the comoving window,

$$\omega_{w0} = \omega_0 - c k_0 v_g = \frac{\omega_p^2}{\omega_0^2}\omega_0 = \frac{\omega_p}{\gamma_g} \qquad (9)$$

which undergoes heavy down-shift due to the classical Doppler-effect. The envelope function $\hat{A}(\xi,t)$ contains not only amplitude modulations but also phase corrections produced by spatiotemporal frequency and wave-number modulations. Therefore, to decompose phase and amplitude contributions, we substitute the phasor presentation of $\hat{A}$ in (8), i.e. substitute $\hat{A}(\xi,t) = a_A(\xi,t)e^{i\vartheta(\xi,t)}$ where $a_A = |\hat{A}|$ and $\vartheta = \arg(\hat{A})$, and from real and imaginary parts we respectively get,

$$[-\frac{1}{\gamma_g^2}(\frac{\partial \vartheta}{\partial \xi})^2 + \frac{1}{c^2}(\frac{\partial \vartheta}{\partial t})^2 - 2\frac{v_g}{c}\frac{\partial \vartheta}{\partial t}\frac{\partial \vartheta}{\partial \xi} - 2\frac{\omega_0}{c^2}\frac{\partial \vartheta}{\partial t}]a_A =$$
$$[\frac{\Omega_p^2 - \omega_p^2}{c^2} - \frac{1}{\gamma_g^2}\frac{\partial^2}{\partial \xi^2} + \frac{1}{c^2}\frac{\partial^2}{\partial t^2} - \frac{2v_g}{c}\frac{\partial^2}{\partial \xi \partial t}]a_A \quad ,(10a)$$

$$2\left(\frac{1}{\gamma_g^2}\frac{\partial \vartheta}{\partial \xi} + \frac{v_g}{c}\frac{\partial \vartheta}{\partial t}\right)\frac{\partial a_A}{\partial \xi} + 2\left(\frac{\omega_0}{c^2} + \frac{v_g}{c}\frac{\partial \vartheta}{\partial \xi} - \frac{1}{c^2}\frac{\partial \vartheta}{\partial t}\right)\frac{\partial a_A}{\partial t} =$$
$$-[\frac{1}{\gamma_g^2}\frac{\partial^2 \vartheta}{\partial \xi^2} - \frac{1}{c^2}\frac{\partial^2 \vartheta}{\partial t^2} + \frac{2v_g}{c}\frac{\partial^2 \vartheta}{\partial t \partial \xi}]a_A \quad (10b)$$

which present coupled equation for phase and amplitude evolutions. The variables $\vartheta$ and $a_A$ are imposed to the following trivial initial conditions,

$$a_A(\xi,t=0) = \hat{A}_0(\xi) \quad , \quad \vartheta(\xi,t=0) = 0 \tag{11}$$

where $\hat{A}_0(\xi) = \hat{A}(\xi,t=0)$ is the initial pulse envelope. The physical significance of the phase factor, $\vartheta$, is that $\varpi = -\partial \vartheta / \partial t$ and $\bar{k} = \partial \vartheta / \partial \xi$ represent respectively local frequency and wave-number corrections (modulations) in the PCMW, appeared due to strong density modulations.

In applying SEA, it should be noticed that rewriting the phase factor $k_0 x - \omega_0 t \equiv k_0 \xi - \omega_{w0} t$ in terms of $\xi$ suggests choosing the carrier frequency in PCMW, $\omega_{w0}$, as the intrinsic fast rate scale, by which the slowness of evolutions should be referenced, i.e. we compare the normalized evolution rates of $a_A$ and $\vartheta$ with the intrinsic carrier frequency in PCMW, $\omega_{w0}$. This is contrary to the previous studies concerning ultra-short pulses, which use the plasma frequency, $\omega_p$, instead of $\omega_{w0}$. As will be seen, this choice leads to substantial extension in the validity scope of SEA regarding the pulse length, as much longer pulses may be subjected to SEA. The validity of our strategy will be firmly demonstrated by showing the consistency of our final results with the implied approximation. We base our calculations on the following general assumptions on the time derivatives of $a_A$ and $\varpi$:

$$\left|\frac{\partial^{n+1} a_A}{\partial t^{n+1}}\right| \ll \omega_{w0} \left|\frac{\partial^n a_A}{\partial t^n}\right|, \qquad \left|\frac{\partial^{n+1} \varpi}{\partial t^{n+1}}\right| \ll \omega_{w0} \left|\frac{\partial^n \varpi}{\partial t^n}\right| \qquad (12)$$

where $n = 0, 1, 2, \ldots$. It is seen that we apply SEA on $\varpi$ instead of $\vartheta$ itself, as $|\partial \vartheta / \partial t| = \varpi$ may reaches values as large as $\omega_{w0}$, due to the present large density modulations. Moreover, it should be strongly emphasized that the above condition

on $\varpi$ (and more generally the SEA) breaks during the crossing of a sharp plasma boundary by the pulse, as it suddenly jumps from $\varpi = 0$ just before entrance to values comparable to $\omega_{w0}$ just after entrance. This phenomenon could happen when the pulse is considered to initiate in the vacuum, outside the plasma, as is usually in real experiments. However, to avoid the tremendous elaboration of calculations, we ignore the transient pulse evolutions during the entrance and consider initialization *inside the plasma*. This was notified because such simplification makes subtleties in determining initial conditions which will be discussed bellow.

Given the conditions (12), we propose a method similar to the usual series-expansion method [48] to solve Eqs. (10a) and (10b). However, the procedure is not as trivial as the well-known examples which are usually treated in this context. To our aim, as in the usual manner, we begin with substitution of the following time-domain Taylor expansion (around $t = 0$) of quantities into the differential equations, (10a) and (10b),

$$a_A(\xi,t) = a_{A.0}(\xi) + t a_{A.t0}(\xi) + t^2 a_{A.tt0}(\xi)/2! + t^3 a_{A.ttt0}(\xi)/3! + .... \quad (13a)$$

$$\Omega_p^2(\xi,t) = \Omega_{p0}^2(\xi) + t\Omega_{pt0}^2(\xi) + t^2\Omega_{ptt0}^2(\xi)/2! + t^3\Omega_{pttt0}^2(\xi)/3! + .... \quad (13b)$$

$$\vartheta(\xi,t) = -t\varpi_0(\xi) - t^2\varpi_{t0}(\xi)/2! - t^3\varpi_{tt0}(\xi)/3! + .... \quad (13c)$$

where (13c) is resulted from integration of $\varpi(\xi,t) = \varpi_0(\xi) + t\varpi_{t0}(\xi) + t^2\varpi_{tt0}(\xi)/2! + ....$ by applying the initial condition (11); 0 and $t$ indices show the initial time and time derivative, respectively. After this substitution, via ordering the result in powers of $t$, we get consecutive algebraic equations among different orders of time derivatives, viz. for $n^{th}$ $t$-power, $n+2^{th}$ time derivative is obtained in terms of $n+1^{th}$ and $n^{th}$ derivatives. Therefore, unknown time derivatives at all levels are found via a forward substitution procedure initiated with the initial conditions including profiles of the initial state and the first time derivative of the solutions; in simple examples this procedure may be summarized into a general recursion relation. The distinctive feature our problem regarding Eqs. (10a) and (10b) is that, *when the pulse is considered to be initiated inside the plasma (as mentioned above, to keep SEA validity)*, the initial time derivatives $a_{A.t0}$ and $\vartheta_{t0}$ cannot be trivially determined (are not given). This is despite the vacuum initialization which leads to $a_{A.t0} = 0$ and $\vartheta_{t0} = 0$ in the cost of SEA breakdown. Fortunately, the SEA conditions (12) provide a way to circumvent this difficulty:

Given the condition (12), depending on our desired accuracy, we ignore the time derivatives beyond a definite order for each of solutions $a_A$ or $\vartheta$. For example let $N$ and $M$ denote the orders of highest significant time-derivatives in $a_A$ or $\vartheta$ expansions, respectively. This implies two approximations $\left|\partial^{N+1}a_A/\partial t^{N+1}\right|\simeq 0$ and $\left|\partial^{M+1}\varpi/\partial t^{M+1}\right|\simeq 0$ which can respectively replace the two unknown initial conditions $a_{A.t0}$ and $\vartheta_{t0}$. Moreover, based on these conditions, we may find the lower order derivatives using a backward substitution procedure, up to finding $a_{A.t0}$ and $\vartheta_{t0}$; in the hierarchy of equations obtained via series-expansion, using the topmost equations we eliminate $\partial^N a_A/\partial t^N$ and $\partial^M \varpi/\partial t^M$ in terms of $\partial^{N-1}a_A/\partial t^{N-1}$ and $\partial^{M-1}\varpi/\partial t^{M-1}$, then, using the subsequent equation, $\partial^{N-1}a_A/\partial t^{N-1}$ and $\partial^{M-1}\varpi/\partial t^{M-1}$ are found in terms of $\partial^{N-2}a_A/\partial t^{N-2}$ and $\partial^{M-2}\varpi/\partial t^{M-2}$, and this elimination continues until we get $a_{A.t0}$ and $\vartheta_{t0}$ in terms of $a_{A0}$ and $\vartheta_0$ (given initial conditions in (11)). Afterward, we may use the usual forward substitution to find the full solution of Eqs. (10a) and (10b) with the desired accuracy. Below we exemplify this procedure, for the first two pairs of the series-expansion hierarchy:

Upon substitution of (13a)-(13c) into (10a) and (10b) from factorizing $t^0$ multipliers we get,

$$[\varpi_0^2 + 2\omega_0\varpi_0]\hat{A}_0 = [\Omega_{p0}^2 - \omega_p^2 - \frac{c^2}{\gamma_g^2}\frac{d^2}{d\xi^2}]\hat{A}_0 + a_{A.tt0} - 2cv_g\frac{da_{A.t0}}{d\xi}, \qquad (14a)$$

$$-cv_g\varpi_0\frac{d\hat{A}_0}{d\xi} + (\omega_0+\varpi_0)a_{A.t0} + \hat{A}_0(\frac{\varpi_{t0}}{2} - cv_g\frac{d\varpi_0}{d\xi}) = 0. \qquad (14b)$$

From $t^1$ factorization we obtain,

$$[2\varpi_0\varpi_{t0} - 2cv_g\varpi_0\frac{d\varpi_0}{d\xi} + 2\omega_0\varpi_{t0}]\hat{A}_0 +$$

$$[\varpi_0^2 + 2\omega_0\varpi_0 - \Omega_{p0}^2 + \omega_p^2]a_{A.t0} = -\frac{c^2}{\gamma_g^2}\frac{d^2 a_{A.t0}}{d\xi^2} - 2cv_g\frac{da_{A.tt0}}{d\xi} + \Omega_{p.t0}^2\hat{A}_0 + a_{A.ttt0}$$ ,(15a)

$$2[\omega_0 + \varpi_0]a_{A.tt0} - 2[cv_g\frac{d\varpi_0}{d\xi} - \varpi_{t0} + cv_g\varpi_0\frac{d}{d\xi}]a_{A.t0} - 2[\frac{c^2}{\gamma_g^2}\frac{d\varpi_0}{d\xi} + cv_g\varpi_{t0}]\frac{d\hat{A}_0}{d\xi}$$

$$= [\frac{c^2}{\gamma_g^2}\frac{d^2\varpi_0}{d\xi^2} + 2cv_g\frac{d\varpi_{t0}}{d\xi} - \varpi_{tt0}]\hat{A}_0 + [2cv_g\frac{d\varpi_0}{d\xi} - \varpi_{t0}]a_{A.t0}$$ . (15b)

Now, according to the general strategy mentioned above, we may ignore $a_{A.ttt0}$ and $\varpi_{tt0}$ respectively in Eqs. (15a) and (15b). Afterward, from resulting equations we may express $a_{A.tt0}$ and $\varpi_{t0}$ in terms of $a_{A.t0}$ and $\varpi_0$, and substitute the results into Eqs. (14a) and (14b) to obtain the latter quantities in terms of $\hat{A}_0$. By noting the conditions (12), we may obtain the sufficient conditions for validity of applied approximations in Eqs. (15a) and (15b) (elimination of $a_{A.ttt0}$ and $\varpi_{tt0}$); it is easily seen that it suffices that $\partial/\partial\xi$ behaves as $|\partial X/\partial\xi| \geq \omega_{w0}|X|/c$ where $X = \{a_{A.tt0}, \varpi_{t0}\}$, which in turn needs the pulse length satisfy,

$$L_p \leq c/\omega_{w0} = \gamma_g c/\omega_p \qquad (16)$$

This allows for very longer pulse lengths in applying SEA, with respect to the condition $L_p \sim c/\omega_p$ (ultra-short pulse length) previously considered in this context [43, 44, 46, 47].

By considering conditions (12) and (16), we may further simplify Eqs. (14) and (15) in order to characterize the most important pulse evolutions, including the amplitude modulation rate $a_{A.t0}$, the frequency corrections (modulations) $\varpi_0$, the spatial frequency-chirp $t\partial \varpi_0 / \partial \xi$, and the temporal frequency-chirp $\varpi_{t0} = \partial \varpi / \partial t$. In the case of (15a), after substitution of the second bracket in its left hand side via Eq. (14a), and keeping the most significant terms, we obtain, $[-2cv_g\varpi_0(d\varpi_0/d\xi) + 2\omega_0\varpi_{t0}]\hat{A}_0 = \Omega_{p.t0}^2 \hat{A}_0$ giving the temporal frequency-chirp,

$$\varpi_{t0} = \frac{cv_g\varpi_0}{\omega_0}\frac{\partial \varpi_0}{\partial \xi} + \frac{\Omega_{p.t0}^2}{2\omega_0}, \qquad (17)$$

which states that this quantity depends not only to plasma-frequency variations, but also to the spatial chirp resulted from induced density modulations. Next, we may use this result in Eq. (15b) to obtain $a_{A.tt0}$ which, together with $\varpi_{t0}$ itself, can be substituted into Eqs. (14a) and (14b) to accomplish a second order accurate (up to $t^2$) solution to $a_{A.t0}$ and $\varpi_0$.

However, in this paper, we prefer to keep the simplicity in calculations of $a_{A.t0}$ and $\varpi_0$, which is done by ignoring $a_{A.tt0}$ and $\varpi_{t0}$ in the right hand sides of (14a) and (14b), respectively. After these simplifications, by combination of two equations, we obtain,

$$\frac{d}{d\xi}[\frac{1}{\omega_0+\varpi_0}\frac{d}{d\xi}(\hat{A}_0\varpi_0)]+\frac{\hat{A}_0}{2c^2v_g^2}[\varpi_0^2+2\omega_0\varpi_0]=\frac{1}{2c^2v_g^2}[\Omega_{p0}^2-\omega_p^2-\frac{c^2}{\gamma_g^2}\frac{d^2}{d\xi^2}]\hat{A}_0 \quad (18a)$$

$$a_{A.t0}=\frac{cv_g}{\omega_0+\varpi_0}\frac{d}{d\xi}(\varpi_0\hat{A}_0). \quad (18b)$$

The ordinary differential equation (18a) may be solved, either numerically or analytically. If we linearize this equation with respect to $\varpi_0/\omega_0 \ll 1$, it takes a familiar form, $(d^2/d\xi^2)[\hat{A}_0\varpi_0]+(\omega_0/cv_g)^2\hat{A}_0\varpi_0 = (\hat{A}_0\omega_0/2c^2v_g^2)[\Omega_{p0}^2-\omega_p^2-c^2\gamma_g^{-2}(d^2/d\xi^2)]\hat{A}_0$ having the following formal solution,

$$\varpi_0=\frac{1}{4\hat{A}_0cv_g}\int_{-\infty}^{0}d\xi'\left\{\sin(\frac{\omega_0}{cv_g}|\xi-\xi'|)[\Omega_{p0}^2(\xi')-\omega_p^2-\frac{c^2}{\gamma_g^2}\frac{d^2}{d\xi'^2}]\hat{A}_0(\xi')\right\} \quad (18c)$$

The above integral may be approximated to get the leading order solution, but a much easier way is simplifying the original equation (18a). This may be done by keeping only the highest order terms, including $2\omega_0\vartheta_{t0}$ and $\Omega_{p0}^2-\omega_p^2$, and finally we obtain,

$$\varpi_0\simeq\frac{\Omega_{p0}^2-\omega_p^2}{2\omega_0}=\frac{\omega_{w0}}{2}(\frac{\Omega_{p0}^2}{\omega_p^2}-1). \quad (19)$$

$a_{A.t0}$ can be obtained by substituting (18c) or (19) into Eq. (18b).

Eq. (19) states that frequency corrections may reach values comparable/ higher than $\omega_{w0}$ in the nonlinear regime where $\Omega_{p0}^2$ substantially differs from $\omega_p^2$. Moreover, putting this equation together with (18b) and (17) we may easily verify that, $a_{A.t0} \ll \omega_{w0}\hat{A}_0$ and $\varpi_{t0} \ll \omega_{w0}\varpi_0$, whence consistency of our formulation regarding the assumptions (12).

Among our final results (17), (18b), (18c) and (19), only the former displays dependence on wakefield evolutions (through $\Omega_{p.t0}^2$). Unfortunately, as will be fully discussed in the next sub-section, wakefield evolutions may not generally be reduced into a closed solution. The exception is ultra-short pulses for which approximations $H_x \simeq 1$ and $J_x \simeq -v_g$ will be verified to hold. However, this fact does not decrease the validity of SEA, as it will be shown that the phase factor is eliminated in the evolutions of $\Omega_p^2$, therefore this quantity evolves as slow as the pulse envelope.

At the end, it is worth mentioning that by substitution of our approximate Eq. (19) into (18b) we find $a_{A.t0} \simeq (c\omega_{w0}/2\omega_0\omega_p^2)(d/d\xi)(\Omega_{p0}^2\hat{A}_0) - (c\omega_{w0}/2\omega_0)d\hat{A}_0/d\xi$. This result is approximately equal to the result given by *Schroeder etal* [44] below their Eq. (5), which in terms of our variables reads as $a_{A.t0}^* \simeq (c\omega_{w0}/2\omega_0\omega_p^2)(d/d\xi)(\Omega_{p0}^2\hat{A}_0)$ where $a_{A.t0}^*$ is measured on the *light speed* commoving window (not *group-velocity* commoving window); the relation between $a_{A.t0}^*$ and $a_{A.t0}$ is $a_{A.t0}^* = a_{A.t0} + c(1-v_g)\partial \hat{A}_0 / \partial \xi$ which verifies coincidence of two results. It should be also mentioned that formulas for spatial frequency chirp (18c) and temporal frequency chirp (17), and explicit relation between frequency-chirp and envelope evolutions are obtained here for the first time. Moreover, fine oscillations at wave number $\omega_0/c$ are predicted to appear in the pulse envelope according to Eqs. (18c) & (18b). These oscillations lead to production of very low-frequency long-wavelength ($k \simeq 0$, $\omega \simeq 0$) modes via beating the carrier wave-number and frequency.

### b. The plasma motion

As previously stated, given the pulse evolutions, the plasma motion is governed by the set of equations (5), (7b) and the Poisson equation (1d), viz. every plasma quantity may be expressed in terms of $\phi$, $H_x$ and $J_x$. Regarding these equations, we firstly aim to notify an important property of their solutions. That is, the solution is decomposed into two parts, the dominant part with slow spatial variations and another small part with fine spatial oscillations proportional to the pulse wave-number. Moreover the effect of temporal oscillations is completely isolated in the latter part.

To expand the above statements, let we examine the following trial solutions to the set of Eqs. (1d), (5), (7b),

$$\phi = \phi_s(\xi,t) + \phi_o(\xi,t) \tag{20a}$$

$$H_x = H_{xs}(\xi,t) + H_{xo}(\xi,t) \tag{20b}$$

$$J_x = J_{xs}(\xi,t) + J_{xo}(\xi,t) \tag{20c}$$

where "s" and "o" respectively show secular and oscillating parts used to satisfy $|X_s| \gg |X_o|$. Let first consider the Poisson equation (1d); after substitution of $n_e$ using Eq. (3c) and expanding up to the first order in small oscillatory quantities, the resulting equation may decomposed into two equations for $\phi_s$ and $\phi_o$, as follows,

$$\frac{\partial^2 \phi_s}{\partial \xi^2} = -\frac{\omega_p^2}{c^2}\left[J_{xs}^* \frac{1+a_A^2/2}{2(\phi_s + H_{xs}^*)^2} + \frac{J_{xs}^*}{2} + 1\right], \tag{21a}$$

$$\frac{\partial^2 \phi_o}{\partial \xi^2} = -\frac{\omega_p^2}{c^2}\{J_{xo}^* \frac{1+a_A^2/2}{2(\phi_s + H_{xs}^*)^2} + \frac{J_{xo}^*}{2} + J_{xs}^* \frac{a_A^2 \cos(2\theta_0 + 2\vartheta)}{4(\phi_s + H_{xs}^*)^2}$$
$$- J_{xs}^*(\phi_o + H_{xo}^*)\frac{1+a_A^2(1+\cos(2\theta_0+2\vartheta))/2}{(\phi_s + H_{xs}^*)^3}\} \tag{21b}$$

where $\theta_0 = k_0 x - \omega_{w0} t$. It is seen that secular potential is completely decoupled from the oscillatory quantities as a result of linearization. In addition, the oscillatory potential is driven dominantly by the third term in the right hand of Eq. (21b) which is independent of unknown oscillatory quantities. Therefore, to the leading order, this equation may be easily integrated utilizing different length scales of oscillatory and secular quantities. To understand this statement, consider the identity,

$$\frac{\partial^2}{\partial \xi^2}[X_s \cos(2\theta_0 + 2\vartheta)] = \frac{\partial^2 X_s}{\partial \xi^2}\cos(2\theta_0 + 2\vartheta) - 4\frac{\partial X_s}{\partial \xi}(\bar{k} + k_0)\sin(2\theta_0 + 2\vartheta)$$
$$-2X_s \frac{\partial \bar{k}}{\partial \xi}\sin(2\theta_0 + 2\vartheta) - 4X_s(\bar{k} + k_0)^2 \cos(2\theta_0 + 2\vartheta)$$

where $X_s$ is an arbitrary secular quantity and $\bar{k} \equiv \partial \vartheta / \partial \xi$. Because the inequality $\partial X_s / \partial \xi \ll k_0 X_s$, the most dominant term in the right hand side (RHS) of above expression is the last term which according to the inequality $\bar{k} \ll k_0$ becomes $4X_s k_0^2 \cos(2\theta_0 + 2\vartheta)$ approximately. Therefore, the dominant term (the third term) of the RHS of (21b) may be approximated by $(-1/4k_0^2)(\partial^2 / \partial \xi^2)[(...)_s \cos(2\theta_0 + 2\vartheta)]$ which finally gives the following leading order solution,

$$\phi_o \approx \frac{\omega_p^2}{16c^2 k_0^2} J_{xs}^* \frac{a_A^2 \cos(2\theta_0 + 2\vartheta)}{4(\phi_s + H_{xs}^*)^2}. \qquad (21c)$$

This result, about $\phi$, has been obtained previously by *Esarey etal* [49] (see Eq. (17a) of [49]) for $J_{xs}^* = -1$ and $H_{xs}^* = 1$.

The same analysis may be carried out In the case of Eqs. (5) and (7b). Let take (5); using (3b-3c) we may express the plasma factors (here $1/\gamma_e$ and $v_{ex}$) in terms of $\phi$, $H_x$ and $J_x$, then linearize the results with respect to $\phi_o$, $H_{xo}$ and $J_{xo}$. In this way, for $1/\gamma_e$ we obtain,

$$\frac{1}{\gamma_e} = \frac{1}{\Gamma_s + \Gamma_o} + \Delta_\gamma \qquad (22a)$$

$$\Gamma_s \equiv \frac{1}{2}(\frac{1 + a_A^2/2}{\phi_s + H_{xs}^*} + \phi_s + H_{xs}^*)$$

$$\Gamma_o \equiv \frac{a_A^2 \cos(2\theta_0 + 2\vartheta)}{4(\phi_s + H_{xs}^*)}$$

$$\Delta_\gamma \equiv -\frac{1}{2(\Gamma_s + \Gamma_o)^2}(1 - \frac{\gamma_\perp^2}{(\phi_s + H_{xs})^2})(\phi_o + H_o)$$

For $v_{ex}$ we notice that (3a) may be simplified into $v_{ex} = 1 - (\phi + H_x^*)/\gamma_e$. Therefore, with the aid of (22a) we obtain,

$$v_{ex} = V_{os} + \Delta_v \qquad (22b)$$

$$V_{os} \equiv 1 - \frac{\phi_s + H_{xs}^*}{\Gamma_s + \Gamma_o}$$

$$\Delta_v \equiv -(\phi + H_{xs})\Delta_\gamma - \frac{\phi_o + H_o}{\Gamma_s + \Gamma_o}$$

It should be noticed that the oscillatory quantity, $\Gamma_o$, and the mixed quantity, $V_{os}$, are not small quantities in spite of $\phi_o$, $H_{xo}$ and $J_{xo}$. We substitute (22a) and (22b) into (5) and linearize the result with respect to $\phi_o$, $H_{xo}$ and $J_{xo}$, and finally decompose the resulting equation into secular (almost zero wave-number) and oscillatory (all possible harmonics of $\theta_0$) parts (we should be careful that multiplication of oscillatory plasma quantities generally results into both secular and oscillatory parts), as,

$$\frac{\partial H_{xs}}{\partial t} + c(S[V_{os} + \Delta_v] - v_g)\frac{\partial H_{xs}}{\partial \xi} = -\frac{\partial \phi_s}{\partial t} + \frac{a_A}{2}\frac{\partial a_A}{\partial t} S\left[(\frac{1}{\Gamma_s + \Gamma_o} + \Delta_\gamma)(1 + \cos(2\theta_0 + 2\vartheta))\right]$$

$$-cS\left[(V_{os} + \Delta_v)\frac{\partial H_{xo}}{\partial \xi}\right] + (\omega_{w0} + \varpi)\frac{a_A^2}{2}S\left[\Delta_\gamma \sin(2\theta_0 + 2\vartheta)\right]$$

(23a)

$$\frac{\partial H_{xo}}{\partial t} + cO\left[(V_{os} + \Delta_v - v_g)\frac{\partial H_{xo}}{\partial \xi}\right] + cO[\Delta_v]\frac{\partial H_{xs}}{\partial \xi} = -\frac{\partial \phi_o}{\partial t} + (\omega_{w0} + \varpi)\frac{a_A^2}{2}\frac{\sin(2\theta_0 + 2\vartheta)}{\Gamma_s + \Gamma_o}$$

$$-cO[V_{os}]\frac{\partial H_{xs}}{\partial \xi} + O\left[(\frac{1}{\Gamma_s + \Gamma_o} + \Delta_\gamma)(1 + \cos(2\theta_0 + 2\vartheta))\right]\frac{a_A}{2}\frac{\partial a_A}{\partial t}$$

$$+(\omega_{w0} + \varpi)\frac{a_A^2}{2}O\left[\Delta_\gamma \sin(2\theta_0 + 2\vartheta)\right]$$

(23b)

where $S[...]$ and $O[...]$ respectively mean secular and oscillatory values of included expressions. In the right hand sides of above equations, we have also used the fact that $S[\sin(2\theta_0 + 2\vartheta)/(\Gamma_s + \Gamma_o)] = 0$ (multiplication of sin and functional of cos does not produce secular parts). In addition we have used the identity

$A_y(\partial A_y/\partial t) = (\omega_{w0} + \varpi)(a_A^2/2)\sin(2\theta_0 + 2\vartheta) + (a_A/2)(\partial a_A/\partial t)[1 + \cos(2\theta_0 + 2\vartheta)]$

Eq. (23b) may be dramatically simplified when we consider the order of magnitude of its different terms; $\partial H_{xo}/\partial t \sim \omega_{w0} H_{xo}$, $\partial H_{xo}/\partial \xi \sim k_0 H_{xo}$, $v_g \partial H_{xs}/\partial \xi \sim a_A \partial a_A/\partial t$ (according to (23a)) and $\partial \phi_o/\partial t \sim (\omega_{w0}/\gamma_g^2) a_A^2$ (according to (21c)). Therefore, to the leading order, Eq. (23b) simplifies to,

$$c(V_{os} - v_g)\frac{\partial H_{xo}}{\partial \xi} = (\omega_{w0} + \varpi)\frac{a_A^2}{2}\frac{\sin(2\theta_0 + 2\vartheta)}{\Gamma_s + \Gamma_o}. \quad (24a)$$

This result may be further simplified by using the definition $V_{os} \equiv 1 - (\phi_s + H_{xs}^*)/(\Gamma_s + \Gamma_o)$ (see (22b)) and the approximation $1 - v_g \approx 1/2\gamma_g^2$. Upon substitution of these equations into (24a) and truncating terms proportional to $\gamma_g^{-2}$, the integration of Eq. (24a) (in the same manner as done for $\phi_o$) ultimately gives,

$$H_{xo} \approx \frac{\omega_{w0} + \varpi}{4k_0 c(\phi_s + H_{xs})} a_A^2 \cos(2\theta_0 + 2\vartheta). \quad (24b)$$

Using the above result we may simplify Eq. (23a), as, we find,

$$S\left[(V_{os} + \Delta_v)\frac{\partial H_{xo}}{\partial \xi}\right] = 0, \quad S\left[\Delta_\gamma \sin(2\theta_0 + 2\vartheta)\right] = 0, \quad (25a)$$

therefore, we have,

$$\frac{\partial H_{xs}}{\partial t} + c(S[V_{os} + \Delta_v] - v_g)\frac{\partial H_{xs}}{\partial \xi} = -\frac{\partial \phi_s}{\partial t} + \frac{a_A}{2}\frac{\partial a_A}{\partial t} S\left[(\frac{1}{\Gamma_s + \Gamma_o} + \Delta_\gamma)(1 + \cos(2\theta_0 + 2\vartheta))\right]$$
.(25b)

Let consider the solution of Eq. (25b); to the leading order we may ignore $\Delta_\gamma$ versus $1/(\Gamma_s + \Gamma_o)$ and $\Delta_v$ versus $v_g$, and we are left with calculation of secular parts $S[\ldots]$, which may be best done by performing integration of the operand over the spatial interval $0 \leq \xi \leq \pi/(k_0 + \bar{k})$ to obtain the average value. For example for $S[(1+\cos(2\theta_0 + 2\vartheta))/(\Gamma_s + \Gamma_o)]$ we have,

$$S\left[\frac{1+\cos(2\theta_0 + 2\vartheta)}{\Gamma_s + \Gamma_o}\right] = \frac{k_0 + \bar{k}}{\pi} \int_0^{\pi/(k_0+\bar{k})} d\xi \left[\frac{1+\cos(2\theta_0 + 2\vartheta)}{\Gamma_s + \Gamma_o}\right] \qquad (26a)$$

It is clear that above integrals eliminates oscillatory harmonics of $2\theta_0 + 2\vartheta$ in all orders except for the zero non-oscillatory one. In addition, the non-oscillatory part remains unaffected due to its very smooth variations. Upon substitution of $\Gamma_s$ and $\Gamma_o$ from (22a) and considering $\xi$ argument in secular quantities as a parameter, the integration may be performed analytically, and the final result is,

$$S\left[\frac{1+\cos(2\theta_0 + 2\vartheta)}{\Gamma_s + \Gamma_o}\right] = \frac{1}{\Gamma_s b}\left(1 - \sqrt{\frac{1-b}{1+b}}\right) \qquad (26b)$$

where $b \equiv (a_A^2/2)/(1 + a_A^2/2 + (H_{xs} + \phi_s)^2)$ has been defined. In the same way we find $S[V_{os}]$, that is,

$$S[V_{os}] = 1 - \frac{\phi_s + H_{xs}}{\Gamma_s \sqrt{1-b^2}} \qquad (26c)$$

Upon substitution of (26b) and (26c) into (25b), we are left with an equation describing evolutions of $H_{xs}$ in terms of variation rates of $\phi_s$ and $a_A$ as driving terms,

$$\frac{\partial H_{xs}}{\partial t} + c(1 - \frac{\phi_s + H_{xs}}{\Gamma_s \sqrt{1-b^2}} - v_g)\frac{\partial H_{xs}}{\partial \xi} = -\frac{\partial \phi_s}{\partial t} + \frac{1}{\Gamma_s}\left(\frac{1}{b} - \frac{1}{b}\sqrt{\frac{1-b}{1+b}}\right)\frac{a_A}{2}\frac{\partial a_A}{\partial t} \quad (27)$$

To understand the behavior of above equation let first consider the so-called adiabatic regime, occurred at very smooth pulses with lengths $L_p \gg c/\omega_p$, for which we have the approximate solution of Poisson equation (21a) in the form $\phi_s + H_{xs} \simeq \gamma_{\perp,s} \equiv \sqrt{1 + a_A^2/2}$. This result is obtained by approximating the left hand of (21a) by zero (see descriptions above Eq. (10) in [45]). Substituting this result into (22a) and the definition for $b$ in (26b), we get $\Gamma_s = \gamma_{\perp,s}$ and $b \equiv (a_A^2/4)/(1 + a_A^2/2)$. It is seen that $b$ is a monotonic function of $a_A$, which reaches $b \to 0.5$ at very large intensities. Therefore, it is a quite good approximation to suppose $b^2 \ll 1$. By applying the above arguments, we may approximate the multiplier of $\partial H_{xs}/\partial \xi$ in (27) by $cv_g$ and obtain the following simplification of (27) at adiabatic conditions,

$$\frac{\partial H_{xs}}{\partial t} - cv_g\frac{\partial H_{xs}}{\partial \xi} = -\frac{\partial \phi_s}{\partial t} + \frac{1}{\Gamma_s b}\left(1 - \sqrt{\frac{1-b}{1+b}}\right)\frac{a_A}{2}\frac{\partial a_A}{\partial t} \quad (28)$$

In fact, the above equation is a very good approximation even for non smooth pulse shapes as long as the intensity is not very high. By noting the relation between time derivatives in transformation from $(x,t)$ to $(\xi,t)$ (from laboratory window (LW) to pulse commoving window (PCMW)), we recall that $\partial H_{xs}(x,t)/\partial t = \partial H_{xs}(\xi,t)/\partial t - v_g \partial H_{xs}(\xi,t)/\partial \xi$, therefore, Eq. (28) is

$$\frac{\partial H_{xs}(x,t)}{\partial t} = -\frac{\partial \phi_s(\xi,t)}{\partial t} + \frac{1}{\Gamma_s(\xi,t)b(\xi,t)}\left(1-\sqrt{\frac{1-b(\xi,t)}{1+b(\xi,t)}}\right)\frac{a_A(\xi,t)}{2}\frac{\partial a_A(\xi,t)}{\partial t} \quad (29a)$$

or,

$$H_{xs}(x,t)-1 = \int_0^t dt'\{-\frac{\partial \phi_s(\xi',t')}{\partial t'} + \frac{1}{\Gamma_s(\xi',t')b(\xi',t')}\left(1-\sqrt{\frac{1-b(\xi',t')}{1+b(\xi',t')}}\right)\frac{a_A(\xi',t')}{2}\frac{\partial a_A(\xi',t')}{\partial t'}\}$$
(29b)

where, in right hand sides, after time derivatives being taken, $\xi$ and $\xi'$ should be evaluated by $\xi = x - v_g t$ and $\xi' = x - v_g t'$.

Eqs. (29a) and (29b), are, in fact, very complex integro-differential equations due to appearance of $\phi_s(\xi',t')$ in the right hand side of these equations, which depends on $H_{xs}(\xi',t')$ itself through the Poisson equation (21a). With this regard, we do not aim to go through the full solution of $H_{xs}$, but rather we want to describe the qualitative behavior of this quantity with the help of obtained equations (29a) and (29b). The latter equation states that, variation in $H_{xs}$ at a given pint $x$ is resulted from cumulative effects of the pulse evolutions during its full passage from that point. Moreover, even after the pulse leaving, effects of pulse evolutions on the front region is transformed back to that point via $\phi_s(\xi',t')$ which transforms the pulse evolution effects backward via Poisson equation. Therefore, we generally expect that $H_{xs}-1$ increases across the pulse, from its front to back, a behavior which is seen in our simulations.

When the adiabatic regime be considered, we would have $\partial \phi_s / \partial t \simeq (a_A / 2\gamma_{\perp,s})\partial a_A / \partial t$ the second term in the integrand of (29b) is highly compensated by $\partial \phi_s / \partial t$, thus the evolution rate is very small and needs very large pulse length to amount in a considerable value for $H_{xs}-1$. Therefore, for short pulses, $H_{xs}$ may be well approximated by unity.

For future uses, we describe the behavior of $\Omega_p^2 = \omega_p^2 n_e / n_{e0}\gamma_e$ (appears in pulse evolutions) in terms of $H_x$, $J_x$, $\phi$. To do this, by substitution of (20a)-(20c) into (3b) and (3a), and making linear expansions in results, we get,

$$\Omega_p^2 = -\frac{J_x}{H_x + \phi} \simeq \Omega_{p.s}^2 + \Omega_{p.s}^2 \left( \frac{J_{xo}}{J_{xs}} - \frac{H_{xo} + \phi_0}{H_{xs} + \phi_s} \right) \tag{30}$$

where $\Omega_{p.s}^2 = -J_{xs} / (H_{xs} + \phi_s)$ This equation shows that despite $n_e$ and $\gamma_e$ themselves, $\Omega_p^2$ contains only very small oscillatory content in the form $\cos(2\theta_0 + 2\vartheta)$. Finally, as an important result verifying our assumptions made in Sec. (III.a), Eq. (30) states that $\Omega_{p.s}^2$ evolves as slow as pulse envelope, as it mimics $H_{xs}$ behavior.

### IV. Stability/ instability against high frequency noise

Here, we consider the presence of a weak background of high frequency (in PCMW) radiations in the plasma and its effect in the overall system dynamics, that is the following general ansatz is examined through the wave equation (1e),

$$A_y(\xi,t) = A_{yq}(\xi,t) + A_{yf}(\xi,t) \qquad (31a)$$

where $A_{yq}$ is the quasi-static solution, discussed in the previous sections, and $A_{yf}$ is a weak, quickly evolving perturbation satisfying $|A_{yf}| \ll |A_{yq}|$. Associated with $A_{yf}$, quickly evolving perturbations are produced in plasma quantities;

$$H_x = H_{xq} + H_{xf} \;,\; J_x = J_{xq} + J_{xf} \;,\; \phi = \phi_q + \phi_f. \qquad (31b)$$

It should be emphasized that analyses in the above sections have been about dynamics of quasi-static quantities ($A_{yq}$, $H_{xq}$, $J_{xq}$ and $\phi_q$) *without using "q" indices on quantities*. When expansions (31a) and (31b) are substituted back to the fundamental equations (1e), (5), (7b) and (1d) subjected to linearization, after some simplifications to be mentioned and subtraction of zero order equations (equations for $H_{xq}$, $J_{xq}$ and $\phi_q$, which discussed in the previous section), it is fairly straightforward to obtain the following equations,

$$\left[\frac{1}{\gamma_g^2}\frac{\partial^2}{\partial \xi^2} - \frac{1}{c^2}\frac{\partial^2}{\partial t^2} + 2\frac{v_g}{c}\frac{\partial^2}{\partial \xi \partial t} - \frac{\Omega_{pq}^2}{c^2}\right]A_{yf} = -\frac{\Omega_{pq}^2 A_{yq}}{c^2 \beta_g}\{\frac{\Omega_{pq}^2}{\omega_p^2}(H_{xf}+\phi_f)+J_{xf}\} \quad (32a)$$

$$\frac{\partial H_{xf}}{\partial t} + c(v_{ex.q}-v_g)\frac{\partial H_{xf}}{\partial \xi} = -\frac{\partial \phi_f}{\partial t} + \frac{A_{yq}}{\gamma_{eq}}\frac{\partial A_{yf}}{\partial t} \quad (32b)$$

$$\frac{\partial J_{xf}}{\partial t} + c(v_{ex.q}-v_g)\frac{\partial J_{xf}}{\partial \xi} = -c\frac{\Omega_{pq}^2}{\omega_p^2 \gamma_{ex.q}^2}\frac{\partial H_{xf}}{\partial \xi} - \frac{\Omega_{pq}^2 v_{exq}}{\omega_p^2 \gamma_{eq}} A_{yq}\frac{\partial A_{yf}}{\partial t} \quad (32c)$$

$$\frac{\partial^2 \phi_f}{\partial \xi^2} = -\frac{\Omega_{pq}^2}{\beta_g c^2}(\frac{\Omega_{pq}\gamma_{\perp q}}{\omega_p})^2 \left(\frac{\Omega_{pq}^2}{\beta_g \omega_p^2}(H_{xf}+\phi_f) - \frac{A_{yq}A_{yf}}{\gamma_{\perp q}^2}\right) - \frac{\omega_p^2}{2c^2}\left(1+(\frac{\Omega_{pq}\gamma_{\perp q}}{\beta_g \omega_p^2})^2\right)J_{xf}$$

(32d)

where we have used the definition $\Omega_p^2 \equiv \omega_p^2 n_e/\gamma_e$. Moreover, in deriving Eqs. (32b) and (32c), it has been noticed that the dominant, secular parts of $H_{xq}$, $J_{xq}$ vary very smoothly over the space, according to the analyses presented in the previous section, viz. according to (27) $\partial H_{xq}/\partial \xi \sim \partial a_A/\partial t$. On the other hand, rapidly-varying (over the space) oscillatory parts of $H_{xq}$, $J_{xq}$ are very small, (see (24b)) and may be ignored when multiplied by fast perturbations. In this regard, for example in Eq. (32b), we have ignored $v_{ex.f}(\partial/\partial \xi)H_{xq}$ versus $v_{ex.q}(\partial/\partial \xi)H_{xf}$. In addition, we have ignored $\gamma_{eq}^{-1}A_{yf}(\partial/\partial t)A_{yq}$ and $(\gamma_e^{-1})_f A_{yq}(\partial/\partial t)A_{yq}$ versus $\gamma_{eq}^{-1}A_{yq}(\partial/\partial t)A_{yf}$ because $A_{yq}$ evolves much more slower than $A_{yf}$. The same simplifications have been made in Eq. (32c). Furthermore, as high rate evolutions are intended here, we may totally ignore the ignored spatiotemporal evolutions of $H_{xq}$ and $J_{xq}$, and set $H_{xq} \simeq 1$ and $J_{xq} \simeq -\beta_g$.

By properly combining the set of Eqs. (32b-32d), we may recover wave-form equations for plasma quantities $H_{xf}$, $J_{xf}$ and $\phi_f$. To do this, we firstly take the time derivative of Eq. (32d) and substitute $\partial(H_{xf} + \phi_f)/\partial t$ and $\partial J_{xf}/\partial t$ respectively from Eqs. (32b) and (32c). Doing so, after some mathematical manipulations we end up with $(\partial^2/\partial \xi^2)(\partial \phi_f/\partial t) = -(\omega_p^2/c)(\partial J_{xf}/\partial \xi)$ which after integration over the space gives,

$$\frac{\partial^2 \phi_f}{\partial \xi \partial t} = -\frac{\omega_p^2}{c} J_{xf}. \tag{33a}$$

Now, we take $\partial/\partial \xi$ of Eq. (32b) and use the above equation in the result to obtain,

$$\frac{\partial^2 H_{xf}}{\partial \xi \partial t} + c\frac{\partial}{\partial \xi}[(v_{xq} - v_g)\frac{\partial H_{xf}}{\partial \xi}] = \frac{\omega_p^2}{c} J_{xf} + \frac{\partial}{\partial \xi}[\frac{A_{yq}}{\gamma_{eq}}\frac{\partial A_{yf}}{\partial t}]. \tag{33b}$$

Next, we take once the time derivative of the above equation and another time its space derivative, being left by two equations respectively for $\partial J_{xf}/\partial t$ and $\partial J_{xf}/\partial \xi$. These two equations are substituted into Eq. (32c), and finally we get,

$$\frac{\partial^3 H_{xf}}{\partial \xi \partial t^2} + c\frac{\partial^2}{\partial \xi \partial t}[(v_{xq} - v_g)\frac{\partial H_{xf}}{\partial \xi}] + c(v_{xq} - v_g)\frac{\partial^3 H_{xf}}{\partial \xi^2 \partial t} +$$

$$c^2(v_{xq} - v_g)\frac{\partial^2}{\partial \xi^2}[(v_{xq} - v_g)\frac{\partial H_{xf}}{\partial \xi}] + \frac{\Omega_{pq}^2}{\gamma_{xq}^2}\frac{\partial H_{xf}}{\partial \xi} = \qquad (33c)$$

$$-\frac{\Omega_{pq}^2}{c}\frac{v_{xq}}{\gamma_{eq}}A_{yq}\frac{\partial A_{yf}}{\partial t} + c(v_{xq} - v_g)\frac{\partial^2}{\partial \xi^2}[\frac{A_{yq}}{\gamma_{eq}}\frac{\partial A_{yf}}{\partial t}] + \frac{\partial^2}{\partial \xi \partial t}[\frac{A_{yq}}{\gamma_{eq}}\frac{\partial A_{yf}}{\partial t}]$$

which its left hand side is entirely in terms of $H_{xf}$. Using the relation $\partial v_{ex.q} / \partial t \simeq 0$, we may rewrite the above equation in a much more compact and interpretable form,

$$\left(\frac{d^2}{dt^2} + \nu_H \frac{d}{dt} + \Omega_H^2\right)\frac{\partial H_{xf}}{\partial \xi} = -\frac{\Omega_{pq}^2}{c}\frac{v_{ex.q}A_{yq}}{\gamma_{eq}}\frac{\partial A_{yf}}{\partial t} + \frac{d}{dt}\frac{\partial}{\partial \xi}[\frac{A_{yq}}{\gamma_{eq}}\frac{\partial A_{yf}}{\partial t}] \quad (33d)$$

$$\nu_H \equiv c\partial v_{ex.q} / \partial \xi, \qquad \Omega_H^2 \equiv \Omega_{pq}^2 / \gamma_{xq}^2 + (v_{ex.q} - v_g)\partial^2 v_{ex.q} / \partial \xi^2$$

which is the equation of the damped driven oscillator with position dependent parameters. The complete time derivative is $d/dt = \partial/\partial t + (v_{ex.q} - v_g)\partial/\partial \xi$ as used before, stating that variations are measured along the fluid element trajectory in the Galilean PCMW, i.e. along

$$\xi = \xi_0 + \int_0^t dt'[v_{ex.q} - v_g]. \qquad (33e)$$

We may obtain a similar equation as (33d) for $J_{xf}$, via the same procedure. We give only the final result,

$$\left(\frac{d^2}{dt^2} + v_J \frac{d}{dt} + \Omega_J^2\right) J_{xf} = -v_J \frac{\Omega_{pq}^2}{\omega_p^2 \gamma_{eq}} A_{yq} \frac{\partial A_{yf}}{\partial t}$$
$$- c \frac{\Omega_J^2}{\omega_p^2} \frac{\partial}{\partial \xi}[\frac{A_{yq}}{\gamma_{eq}} \frac{\partial A_{yf}}{\partial t}] - \frac{d}{dt}[\frac{\Omega_{pq}^2 v_{xq}}{\omega_p^2 \gamma_{eq}} A_{yq} \frac{\partial A_{yf}}{\partial t}]$$
(34a)

$$v_J \equiv c\partial v_{ex.q}/\partial \xi - (\gamma_{xq}^2/\Omega_{pq}^2)(v_{ex.q} - v_g)(\partial/\partial \xi)(\Omega_{pq}^2/\gamma_{xq}^2), \quad \Omega_J^2 \equiv \Omega_{pq}^2/\gamma_{xq}^2$$

which according to (33a) gives also,

$$\left(\frac{d^2}{dt^2} + v_J \frac{d}{dt} + \Omega_J^2\right) \frac{\partial^2 \phi_f}{\partial \xi \partial t} = v_J \frac{\Omega_{pq}^2}{c\gamma_{eq}} A_{yq} \frac{\partial A_{yf}}{\partial t}$$
$$+ \Omega_J^2 \frac{\partial}{\partial \xi}[\frac{A_{yq}}{\gamma_{eq}} \frac{\partial A_{yf}}{\partial t}] + \frac{d}{dt}[\frac{\Omega_{pq}^2 v_{xq}}{c\gamma_{eq}} A_{yq} \frac{\partial A_{yf}}{\partial t}]$$
(34b)

The set of Eqs. (32a), (33d), (34a) and (34b) present a system of coupled spatiotemporal oscillators, which fully describe the laser plasma evolutions beyond the quasi-static regime. At very low laser intensities, these equations may be easily combined to recover the well-known quasi-linear equations for momentum and density, suited in common studies of Raman scatterings, viz. at very low intensities we would have $\gamma_{eq} \simeq 1$, $v_{ex.q} \simeq 0$, $\{v_H, v_J\} \simeq 0$, $\Omega_H^2 \simeq \Omega_J^2 \simeq \omega_p^2$, $H_{xf} \approx \gamma_{ef} - \beta_g p_{ex.f} - \phi_f \simeq A_{yq} A_{yf} - \beta_g p_{ex.f} - \phi_f$ and $J_{xf} \simeq p_{ex.f} - \beta_g n_{ef}$. The outcome of plasma wave formulation in terms of driven simple oscillators is an important benefit of our approach over the previous approaches, which produces a manifest analogy with the quasi-linear regime. This simplicity attained because, even in the presence of the strong plasma wave, the unperturbed profiles of $H_x$ and $J_x$ ($H_{x.q}$ and $J_{x.q}$) remain approximately unmodulated (spatial constants); if we have used the density presentation for plasma wave (plasma wave in terms of $n_e$), as in the previous studies, the unperturbed density profile ($n_{e.q}$) would be modulated in the presence of wakefield, and the form of the plasma wave would become very complicated. This is while within our system of coupled oscillator equations, (32a), (33d), (34a) and (34b), we may immediately deduce the possibility of parametric self-amplification according to analogy with the quasi-linear regime; in the linear regime this self-amplification leads to parametric instabilities (exponentiations). Furthermore, as their formulations are similar in nature, we may even understand the main distinguishing points and signatures of quasi-linear and non-linear regimes:

In the so-called adiabatic regime of smooth pulse interactions, according to the analyses presented in the previous section (note that in that section all quantities have been quasi-static without being indexed by "q") we have for the secular part of $v_{ex.q} \ll 1$, and the secular parts of plasma wave parameters reduce to,

$$\Omega_H^2\big|_{ad} \simeq \Omega_J^2\big|_{ad} \simeq \Omega_{pq}^2\big|_{ad} \simeq \frac{\omega_p^2}{\sqrt{1+a_{Aq}^2}}, \; v_H\big|_{ad} \simeq 0, \; v_J\big|_{ad} \simeq 0 \qquad (35)$$

At these conditions Eqs. (33d), (34a) and (34b) become very simplified, and quite analogous to the quasi-linear regime. The only superior effect is the well-known growth rate reduction by the relativistic mass increase [39-42]. Although, we also should add the general effect of the finite pulse length, which leads to additional effects with respect to the infinite-length single-mode pump wave, including spatiotemporal behaviors [17-20] and changing the matching condition between the scattered and pump waves. For example, in backscattering, for the scattered Stokes wave (indexed by "s"), we may have $k_s \simeq -k_0 + \omega_p/c$ and $\omega_s \simeq \omega_0 - \omega_p$ instead of $k_s \simeq -k_0$ and $\omega_s \simeq \omega_0 - \omega_p$ (the phase relation is not only on the frequency but on both the frequency and wave number).

Despite the above simplicity attained at smooth pulse shapes, the situation may become complicated when the pulse shape initially has a quickly rising part. At these conditions, due to the wakefield excitation, the local frequencies $\Omega_H^2$, $\Omega_J^2$ and $\Omega_p^2$ (see Eqs. (33d), (34a) and (34b)) becomes both highly modulated and displaced with respect to each other, and dissipation factors $\nu_H$ and $\nu_J$ becomes non-zero. Moreover, as is seen in (33e), the trajectory of each fluid element deviates highly form $\xi = \xi_0 - v_g t$. In these regards, and in terms of quantities $H_x$ and $J_x$, one may imagine the plasma disturbance as a collection of transverse oscillators whose centers perform longitudinal oscillations and whose frequencies change with time. As a result a complex phase relation is produced between the oscillators and both the pump and the scattered electromagnetic waves. These behaviors leads at most into two consequences; (i) broadening in the scattered wave spectrum (ii), possible losing of the plasma wave resonance with driving electromagnetic pump and scattered waves. Moreover, if an resonance be pertained at all, the Stokes (anti-Stokes) wave may display anomalous shifts in $\omega - k$ plan with respect to the pump wave $(k_0, \omega_0)$, compared to the usual predictions, viz. for example, for the back-scattered wave, the relations $k_s + k_0 = \omega_p / c(1 + \langle a_{Aq}^2 \rangle)^{1/4}$ and $\omega_s - \omega_0 = -\omega_p / (1 + \langle a_{Aq}^2 \rangle)^{1/4}$ may be not fulfilled even approximately. In fact, the displacements of sidebands in mode-space are determined by frequencies $\Omega_H^2$, $\Omega_J^2$ at the most effective parts of the plasma wave. All these behaviors are observed in our simulations.

## V. Numerical and simulation investigations

### a. General considerations on simulation methods and setups

We may validate our obtained results, and further investigate the problem under consideration, via direct numerical solution of our basic equations (1a)-(1e) (the fluid-Maxwell system of equations). However, in this way, we encounter the problem of unpredictable numerical stability of the proposed computational scheme. This problem is quite well-known in the context of computational fluid dynamics (CFD) when nonlinear equation-systems like ours are concerned; owing to the nonlinearity, the stability generally depends on the applied physical conditions. Moreover, the stability criteria could not be established analytically [50]. Therefore, it is not usually possible to determine/guarantee the stability in advance and often we need to evaluate this issue case-by-case and after implementation. Here, we use a second order upwind algorithm [50], very similar to what has been implemented in Ref [51]. Regarding the numerical instability, we encounter limitations over the applied physical parameters. For example, for a given intensity, when we increase the pulse length we observe appearance of negative density peaks (a common sign of crash in CFD) which rapidly grow and eventually destabilize the system (see Ref. [52]). We do not aim to apply much-elaborated computational fluid schemes (like WENO [50]) which may offer more extended domain of numerical-stability, rather we confine ourselves to the applicable parameter space, and use complementary particle-in-cell (PIC) simulations over our uncovered parameter space.

The PIC method, on the other hand, suffers from the so-called finite-grid instability (see [53] and references therein). That is the numerical noise (disorders in electromagnetic fields and particle distributions in the phase-space) cannot be totally switched off (up to the round-off errors) when desired, i.e. even if an initially noise-free numerical setup be chosen, the noise is rapidly produced to a level determined by the finite numerical resolution of the simulation instant. This is why we may not define a plasma with initial zero temperature in a PIC simulation, i.e. the temperature rapidly grows to a level defined by the inherent noise in particle velocities [53]. Regarding, these discrepancies, we may not examine our system in the absence of noise (as is desired for examining the ideas proposed in Sec. IV) in the context of PIC simulations. This is despite the fluid solution method, which allows for total noise elimination up to the round-off errors. Therefore, whenever stable, the fluid solution may be applied to demonstrate the effect of noise (as proposed in Sec. IV) by comparing its results in the presence and absence of the initial noise. On the other hand, as real experimental systems, always include noise content, we may use PIC simulations for parametric studies of pulse evolutions. Moreover, we may go beyond the cold fluid model summarized in Eqs. (1a)-(1e) and investigate the kinetic effects (non-zero temperature effects), by comparing results from cold fluid and PIC simulations at non-zero temperatures.

We have conducted several instances of PIC (for simulation code see [53, 54]) and cold-fluid simulations, at different pulse intensities, lengths and shapes, and plasma densities –some of them are presented here. For all presented simulations, the laser pulse is launched into a slab quiescent plasma from its left boundary, and we invariably set $\lambda = 1\mu m$, $n_{e0}/n_c = 0.01$ and $k_B T_{e0} = 48.5 \text{eV}$ where $\lambda$, $n_{e0}$, $n_c$, $T_{e0}$ and $k_B$ are respectively the laser wavelength, the initial plasma density, the critical density ($n_c = \varepsilon_0 m_e \omega_0^2 / e^2$), the initial electron temperature (typical in laser plasma interactions) and the Boltzmann constant. This is while the dimensionless pulse amplitude ($a_0$), duration (denoted by $\tau_L$) and pulse shape (denoted by $SF$) vary among different runs. The pulse shape comprises three rise, flat and fall parts into a trapezoid shape whose straight sides are replaced by sinusoidal curves. For each value of this factor, we give an ordered triple array as $SF = [rise \text{ time}, flat \text{ time}, fall \text{ time}]$. Totally, we discuss three values of $a_0$, $a_0 = \{0.5, 1, 2\}$, two values of $\tau_L$, $\tau_L = \{200, 300\}\text{fs}$, and four values of $SF$, $SF = \{[100,0,100],[30,140,30],[150,0,150],[30,240,30]\}\text{fs}$.

In cold fluid simulations we are able to introduce an initial small disturbance in the density of the quiescent plasma (denoted by $\delta n_{nois}$), acting as an artificial noise which its level (denoted $\delta n_{nois0}$) may be controlled from an input parameter. Here we set either $\delta n_{nois0}/n_{e0} = 0$ or the typical value $\delta n_{nois0}/n_{e0} \sim 10^{-4}$, and respectively refer to the corresponding run by "without noise" and "with noise". Though, in the absence of noise, the fluid scheme is stable over the mentioned physical parameters, this stability may be lost for some pulse shapes (as will be mentioned in the next subsection) in the presence of the noise. As will be seen, this important nonlinear behavior is due to interference between physical and numerical instabilities.

Using this set of simulations we are not only able to validate our analytical results on self-modulation, but also we are able to clearly identify the role of initial noise in destabilizing the laser propagation and impact of physical parameters in this complex phenomenon.

### b. Results and discussions

First of all, In Fig.1, we plot the pulse envelope evolutions together with frequency modulations as predicted by our analytical equations, (13a), (18b) and (18c), for pulse parameters $a_0 = 2$, $\tau_L = 200\text{fs}$ and $SF = [100, 0, 100]$. In plotting the envelope profiles, we have substituted $\varpi_0$ from (18c) into (18b) and then substituted the result for $a_{A,t0}$ into (13a) while the terms beyond $t^2$ have been ignored in this equation, i.e. $a_A(\xi,t) \simeq a_{A,0}(\xi) + t a_{A,t0}(\xi)$. It is clearly seen that the modulated frequency acts as an effective potential in driving amplitude evolutions as suggested by Eq. (18b). Also, appearance of fine oscillation in envelope is predicted by Eq. (18c), as mentioned at the end of Sec. III.A.

Next, we make comparison between results of two simulation methods with each other and with presented analyses. In this line, in Fig. 2, we have plotted the fundamental quantities $H_x$, $J_x$ and $A_y$ versus $\xi$ at different interaction times $t = 300\text{fs}$ and $t = 400\text{fs}$, from both fluid "with noise" and PIC simulations, at pulse parameters $a_0 = 2$, $\tau_L = 200\text{fs}$ and $SF = [100, 0, 100]$. In addition, to be verified, we have plotted our analytical results for $A_y$ envelope in the manner described just above Fig. 1. In the case of $H_x$, PIC results are $x - h_x$ snapshots of electrons where $h_x$, in correspondence with $H_x$, is defined as $h_x \equiv \gamma - \beta_g p_x - \phi(x)$ for each individual electron with position $x$ and longitudinal momentum $p_x$.

As it is seen in Fig. 2, for considered pulse parameters, both PIC and fluid models are in very good agreement about the secular variations of $H_x$ and $J_x$ profiles inside the pulse. This is while, the kinetic model, which definitely includes non-zero temperature effects, predicts amplification in fine oscillations of these profiles across the pulse, an effect which eventually leads to formation of a chaotic pattern in the phase snapshots at sufficiently large distances from the pulse front. This phenomenon may play an important role in the mechanisms of electron trapping in SMLWFA [27]. Despite these superior kinetic effects in plasma motion, pulse evolutions quite matched in both kinetic and fluid models. Moreover, in this case, an excellent agreement is observed among simulations and our analytical results described on Fig. 1.

Fig. 2, also, presents an example of spatiotemporal behavior of $H_x$, which clearly shows the convective nature of evolutions in this quantity, as described by Eq. (29b). Also fine oscillations are observed in fluid results for this quantity as predicted by Eq. (24b).

Now, before going further in pulse evolution analyses, we aim to demonstrate that when smooth pulse shapes (pulses with smooth rise and fall parts) are considered, the system is extremely sensitive to the presence of an initial noise content. In Fig. 3, we compare the action of fluid simulations, after interaction time of 300fs, in the absence and presence of the initial small noise content at two different pulse shapes 1-sharp rise shape $SF = [30,140,30]$ (panels (a)-(c)) and 2-smooth rise shape $SF = [100,0,100]$ (panels (d)-(f)); other pulse parameters are kept identical, i.e. peak amplitude at $a_0 = 2$, and total duration at $\tau_L = 200$fs. The initial noise content is introduced as a small density disturbance in the upstream of the plasma with amplitude of order of $\delta n_{nois0} \sim 10^{-4} n_{e0}$, as is shown on the insets of panels (b) and (e). Surprisingly, despite absolute neutrality of the first setup ($SF = [30,140,30]$) with respect to the imposed small disturbance, the second setup develops very large oscillations in response, in both $H_x$ and $n_e$ quantities. In other words, the smooth pulse propagation may become unstable in the presence of fine irregularities in the initial plasma profile and/or electromagnetic fields. This important result could not be obtained from PIC simulations, because the noise content could not be totally eliminated in this method. However, at latter times, the destabilized oscillations are subjected to numerical instabilities and become further and further amplified in unphysical manner. This makes (our) fluid simulations unsuitable for long term studies of smooth pulses. In the following, we utilize PIC simulation to study the physical instability of pulse propagation, along with keeping this in mind that such instability is, in fact, seeded by the initial noise content of the plasma and fields, viz. according to numerically stable fluid simulations, the instability does not occur if the noise be totally eliminated.

Next, in Figs 4-10, we reveal the essential differences produced by the pulse shape effects in the system behavior in developing parametric instabilities, as viewed in plasma dynamics. In Figs. 4 and 5, from PIC simulations, we respectively summarize the spatiotemporal behaviors of $H_x$ and $J_x$ at laser intensity $a_0 = 2$ and duration $\tau_L = 300\text{fs}$, and for two different pulse shapes $SF = [150, 0, 150]$ (left columns) and $SF = [30, 240, 30]$ (right columns). In Figs. 4 and 5, we observe that while the smooth pulse shape ($SF = [150, 0, 150]$) develops large amplitude quasi-coherent oscillations in both $H_x$ and $J_x$ in the course of time, in its spite, the quickly rising pulse ($SF = [30, 240, 30]$) produces irregular and much weaker perturbations. In Figs. 6 and 7, we respectively present the same plots as Figs. 4 and 5 with the same corresponding pulse shapes but a different intensity $a_0 = 1$. It is seen that compared to $a_0 = 2$, at this lower intensity ($a_0 = 1$) more stronger oscillations appear in both $H_x$ and $J_x$ when the quickly rising pulse shape is applied, while the plasma behavior at smooth pulse shape remains qualitatively similar for both used intensities. We continue to further lowering the laser intensity to reach the sub-relativistic regime by setting $a_0 = 0.5$. In this regard, in Figs. 8 and 9, we present additional plots respectively for $H_x$ and $J_x$ at this sub-relativistic (quasi-linear) intensity, $a_0 = 0.5$, for same pulse shapes as in the previous cases, $a_0 = 2$ and $a_0 = 1$. It is seen that, in the covered region $-50 \leq \xi \leq -20$, by reducing the intensity down to the sub-relativistic value ($a_0 = 0.5$) the plasma behavior (as seen in $H_x$ and $J_x$) is quite altered with respect to relativistic case ($a_0 = 2$), i.e. oscillations are more stronger in the case of quickly rising pulse shape.

In Fig. 10 we show snapshots of $H_x$ over the interaction regions that, in order to produce clear resolution of oscillations, remained uncovered in Figs. 4 and 8. Here, we plot $H_x$ both over the full interaction region and also magnified uncovered region for considered low and high intensities, $a_0 = 2$ and $a_0 = 0.5$, and pulse shapes $SF = [150, 0, 150]$ and $SF = [30, 240, 30]$, at interaction time $t = 300\,\text{fs}$. One may observe that, at relativistic intensities, oscillations finally evolve into chaotic structures, while remain regular at sub-relativistic intensities.

Moreover, as the instantaneous laser field is also plotted inside each panel, one may find local correlations between laser and plasma variations.

The development of large oscillations in $H_x$ and $J_x$ observed both on fluid simulations of Fig. 3 and in present PIC simulations of Figs. 4-10, under the action of smooth and/or lower intensity pulses, may be described in attribution to development of Raman instability as summarized in Sec. IV. In order to verify this idea, we should consider the self-consistent pulse ($A_y$) evolutions in the mode space ($k-\omega$ plane). To this aim, in Figs. 11-14 we summarize $k-\omega$ maps obtained via the time-space Fourier transform of $A_y$ at desired times for different pulse parameters. These plots include both forward (positive $\omega$, positive $k$) and backward (negative $\omega$, positive $k$) going radiation components. At each of our figures, we have three column of panels which correspond to different interaction times, in order from left to right, $t=100\text{fs}$, $t=200\text{fs}$ and $t=300\text{fs}$. In addition, each figure comprises four rows of panels, the upper two rows (panels (a-f)) correspond to the smooth pulse shape and the lower two rows (panels (g-l)) correspond to the quickly rising pulse shape; out of these two pairs of rows, the second (d-f) and fourth (j-l) rows isolate negative-$\omega$ (backward) modes, in order to magnify the amplitude of these modes versus the strong forward-going pump wave. In Figs. 11, 12 and 13 the same parameters as in Figs. 4, 6 and 8, are used respectively, say, they are given for different intensities $a_0 = 2$ (Fig. 11), $a_0 = 1$ (Fig. 12) and $a_0 = 0.5$ (Fig. 13), the identical pulse duration $\tau_L = 300\text{fs}$, the smooth shape $SF = [150, 0, 150]$ (panels (a-f) in each figure) and the quickly rising shape $SF = [30, 240, 30]$ (in panels (g-l) in each figure). Fig. 14 uses the same intensity as Fig. 11 but different pulse duration $\tau_L = 200\text{fs}$ (same as Fig. 3) and pulse shapes $SF = [100, 0, 100]$ (as smooth) and $SF = [30, 140, 30]$ (as quickly rising). As, is seen, the results presented in Figs. 10-14 confirm the occurrence and different behavior of Raman scattering at different pulse parameters, in a close correlation with observed plasma behavior (Figs. 4-10), and in agreement with analyses in presented Sec. IV:

At sub-relativistic intensities (see Figs. 13, 8 and 9 all at $a_0 = 0.5$) the Raman scattering behaves ordinarily, as given by the quasi-linear theory (note $\omega_p = 0.1\omega_0$ for our parameters; see arrows set on figures). As the instability rate is proportional to pump wave amplitude $a_0$, it develops faster when the pulse reaches its peak amplitude more quickly, that is for quickly rising pulse. Therefore, the observed behavior of plasma oscillations (Figs. 8 and 9) as well as the mode space (Fig. 13) is ordinary and quite within the ordinary quasi-linear theory. This is while, when we enter the relativistic regime, as described at the end of Sec. IV, the quick pulse rise leads to new factors like broadening and lose of resonance, which compete with exponentiation. This is why the effect of pulse quick rise-up becomes eventually reversed at high intensities ($a_0 = 2$, see Figs. 11, 4 and 5) and destruct Raman instability formation. Note this is not only due to the well known effect of relativistic mass increase (see Eq. (35) for smooth pulse), as these effect cannot produce order of magnitude or higher differences appeared between different pulse shapes in Fig. 10. When the intensity is relativistic but moderate ($a_0 = 1$, see Figs. 12, 6 and 7) the Raman instability appears but at an anomalous Stokes shift in $k-\omega$ plane. On Fig. 12, for $a_0 = 1$, we observe a shift twice the quasi-linear prediction $k_s - k_0 \simeq 2\omega_p / c$ and $\omega_s + \omega_0 \simeq 2\omega_p$, again in agreement with predictions at the end of Sec. IV. This phenomenon may be understood by noting the electron density structure shown on Fig. 6(f). When we shorten the pulse length at intensity $a_0 = 2$ from $\tau_L = 300\text{fs}$ to $\tau_L = 200\text{fs}$, as is seen in Fig. 14, we observe increase in the spectrum broadening around the Stokes position, again in agreement with analyses in the end of Sec. IV.

At the end of this section, we may notify appearance of very low-frequency long wavelength ($k \simeq 0$, $\omega \simeq 0$) modes in $k-\omega$ plan at high intensity quickly rising pulses, in agreement with descriptions in the end of Sec. III.A.

## VI. Conclusions and remarks

We would like to point out numbers of our conclusions and remarks, regarding the importance of the present work in resolving interesting problems in laser plasma accelerator [2] and nonlinear pulse evolutions [6, 17-24, 43, 44, 46,47]:

1- As formulated in Eqs. (18b) and (18c), we described self-modulation of a short, relativistic polarized laser pulse in underdense plasma, and showed that this phenomenon is dominantly driven by spatial frequency-chirp in the commoving window (see Fig. 1). The frequency-chirp itself is dominantly induced by density modulations in the excited wakefield (see Eq. (19)). Our results are in excellent agreement with presented numerical simulations (see Fig. 2). Generalization of Eqs. (18b) and (18c) to 3D is straightforward; 3D version of the wave equation (8) is obtained by adding $\nabla_\perp^2 \hat{A}$ to the left hand side of the present equation where $\nabla_\perp^2 \equiv \partial^2/\partial y^2 + \partial^2/\partial z^2$ and $\hat{A}$ is now $\hat{A}(\xi,y,z,t)$. It is fairly straightforward to show that this correction changes Eq. (18c) as,

$$\varpi_0 = \frac{1}{4\hat{A}_0 c v_g} \int_{-\infty}^{0} d\xi' \left\{ \sin(\frac{\omega_0}{c v_g}|\xi - \xi'|)[\Omega_{p0}^2(\xi',y,z) - \omega_p^2 - c^2 \nabla_\perp^2 - \frac{c^2}{\gamma_g^2}\frac{\partial^2}{\partial \xi'^2}]\hat{A}_0(\xi',y,z) \right\},$$

and leaves Eq. (18b) unchanged.

2- We have summarized the important features of the plasma dynamics in the self-modulated laser field. Specially, we found that for an ultra-short pulse $H_x$ (electron energy in PCMW) and $J_x$ (electron flux in PCMW) remain approximately constant (see e.g. regions near the pulse front in Figs. 2, 4). In spite, for longer pulses (considered here) this conservation breaks, and reduction in $H_x$ amounts to a noticeable value at sufficiently large distances from the pulse front (see Figs. 4, 10). This phenomenon can play a significant role in electron injection into the acceleration phase (trapped population) in SMLWFA, as it causes the trajectory of wave-body electrons to get close to the separatrix between trapped and untrapped trajectories. In terms of the cold-fluid wave-break threshold, $\phi_{W.B}$, [2] which is obtained by setting $v_{ex} = v_g$ at the maxima of the wake potential [2], we would have,

$$\phi_{W.B} = H_{x.\min} + \gamma_\perp \gamma_g (1 - \beta_g)$$

which predicts reduction of wave-break amplitude across the pulse due to the reduction of $H_x$ (see Figs. 4, 10 and Eq. 29b).

3- Oscillatory and chaotic patterns in $x - h_x$ plots of Figs. 6, 8, 10, produced by interplay between kinetic effects and light scattering, according to descriptions in Sec. IV and Sec. V, also lead to enhanced electron trapping, as before suggested by Ref. [27].

4- Because our self-modulation description predicts envelope breakup (see Figs. 1, 2) in close agreement with simulations, we conclude that, in the presence of strong wakefield at relativistic laser intensities, the pulse breakup is developed mainly by self-modulation. This is despite the quasi-linear regime in which the breakup is developed by forward Raman instability [17]. When the pulse-shape is initially smooth, the wake excitation and self-modulation are seeded by Raman backward instability, leading to two stages breakup. In this case, we did not repeat the simulation results for the second stage, as, some may be found in other literatures e.g. [35].

4- According to Sec. IV and related simulation results in Sec. V, smooth pulse shapes behave ordinarily in relation with Raman scatterings while quickly rising pulses at high intensities behave anomalously, generally speaking. The sidebands may be highly displaced for intense quickly-rising pulses, and the instability becomes almost halted (beyond the relativistic mass increase effect) at sufficiently high intensities (here $a_0 = 2$).

5- By formulation of the plasma wave in terms of $J_x$ and $H_x$, we have presented a way for refining the unstable plasma perturbations and important hidden plasma evolutions, by taking away the highly modulated plasma background (see Figs. 4, 6, 8, 10). Therefore, we have presented a new theoretical ground and data mining methodology for employing simulations in investigation of instabilities and plasma evolutions at complex nonlinear regimes, which may be generalized to 3D cases.

**Figure Captions**

**Figure 1(color online):** pulse envelope, $a_A$, at different times in picoseconds (solid curves), and frequency corrections, $\varpi_0 \hat{A}_0$, (dash-dot curve) for parameters given in text.

**Figure 2 (color online):** $H_x$, $J_x$ and $A_y$ at $t = 100\text{fs}$ (a-c) and $t = 200\text{fs}$ (d-f), for pulse parameters given in the text. Black, green and blue data respectively correspond to PIC simulation, fluid "with noise" simulation and analytical solution (see text).

**Figure 3 (color online):** profiles for $H_x$, $n_e$ and $A_y$ in the absence (green curves) and presence (blue curves) of initial noise content, for pulse shape $SF = [30, 140, 30]$ (left column) and pulse shape $SF = [100, 0, 100]$ (right column). Inset in panel (d) shows the same data as outset but with full range of $H_x$ axis. Insets in (b) and (e) represent the density profile at upstream of plasma.

**Figure 4 (color online):** $x - h_x$ snapshots of electrons at different times for $a_0 = 2$, $\tau_L = 300\text{fs}$ and smooth $SF = [150, 0, 150]$ (left column) and quickly rising $SF = [30, 240, 30]$ (right column) pulse shapes.

**Figure 5 (color online):** $J_x$ profiles for same parameters and settings as fig. 4.

**Figure 6 (color online):** $x - h_x$ snapshots of electrons for intensity $a_0 = 1$. Other parameters and settings are same as those in fig. 4.

**Figure 7 (color online):** $J_x$ profiles for same parameters and settings as fig. 6.

**Figure 8 (color online):** $x - h_x$ snapshots of electrons for intensity $a_0 = 0.5$. Other parameters and settings are same as those in fig. 4.

**Figure 9 (color online):** $J_x$ profiles for same parameters and settings as fig. 8.

**Figure 10 (color online):** $x - h_x$ snapshots of electrons (blue dots) and instantaneous laser field (green curves) at interaction time $t = 300\text{fs}$ for $\tau_L = 300\text{fs}$ and different intensities $a_0 = 2$ (upper two rows) and $a_0 = 0.5$ (lower two rows), and different smooth $SF = [150, 0, 150]$ (left column) and quickly rising $SF = [30, 240, 30]$ (right column) pulse shapes. The second and fourth rows only magnify the data over the selected position range.

**Figure 11 (color online):** $k - \omega$ map of $A_y$ for $a_0 = 2$ and $\tau_L = 300\text{fs}$ at different interaction times, $t = 100\text{fs}$ (first column), $t = 200\text{fs}$ (second column) and $t = 300\text{fs}$ (third column), and for different smooth $SF = [150, 0, 150]$ (upper two rows; a-f) and quickly rising $SF = [30, 240, 30]$ (lower two rows; g-l) pulse shapes. The second and fourth rows only magnify the data plotted in the first and third rows respectively, over the selected $\omega$ range, to appear more clearly on the color presentation. The vertical and horizontal arrows point to important locations on axes (see text).

**Figure 12 (color online):** $k - \omega$ map of $A_y$ for $a_0 = 1$. Other parameters and settings are same as fig. 11.

**Figure 13 (color online):** $k - \omega$ map of $A_y$ for $a_0 = 0.5$. Other parameters and settings are same as fig. 11.

**Figure 14 (color online):** $k - \omega$ map of $A_y$ for same settings and laser intensity as fig. 11 but different pulse duration $\tau_L = 200\text{fs}$ and pulse shapes $SF = [100, 0, 100]$ (as smooth) and $SF = [30, 140, 30]$ (as quickly rising).

## References

[1] G. A. Mourou, T. Tajima, and S. V. Bulanov, **Rev. Mod. Phys. 78**, 309 (2006).

[2] E. Esarey, C. B. Schroeder, and W. P. Leemans, **Rev. Mod. Phys. 81**, 1229 (2009).

[3] R. Kodama etal, **Nature 412**, 798 (2001).

[4] A. G. MacPhee etal, **Phys. Rev. Lett. 104**, 055002 (2010).

[5] A. L. Lei etal, **Phys. Rev. Lett. 96**, 255006 (2006).

[6] W. B. Mori, **IEEE Journal of Quantum Electronics 33**, 1942 (1997).

[7] M. D. Feit, J. C. Garrison, and A. M. Rubenchik, **Phys. Rev. E 53**, 1068 (1996).

[8] L. Willingale etal**, Phys. Rev. Lett. 96**, 245002 (2006).

[9] J. Krall, A. Ting, E. Esarey, and P. Sprangle, **Phys. Rev. E 48**, 2157 (1993).

[10] A. Ting etal, **Phys. Plasmas 4**, 1889 (1997).

[11] G. Q. Liao etal, **Phys. Rev. Lett. 114**, 255001 (2015).

[12] M.I. K. Santala etal, **Phys. Rev. Lett. 84**, 1459 ( 2000).

[13] D. Wu etal, **Nucl. Fusion 57**, 016007 (2017).

[14] L. Willingale etal, **Phys. Rev. Lett. 102**, 125002 (2009).

[15] Y. T. Li etal, **Phys. Rev. E 72**, 066404 (2005).

[16] T. M. Antonsen Jr. *and* P. Mora, **Physics of Fluids B 5**, 1440 (1993).

[17] W. B. Mori, C. D. Decker, D. E. Hinkel, T. Katsouleas, **Phys. Rev. Lett. 72**, 1482 (1994).

[18] N. E. Andreev *and* V. I. Kirsanov, **Phys. Plasmas 2**, 2573 (1995).

[19] C. J. McKinstrie , **Phys. Plasmas 3**, 4683 (1996).

[20] G. Shvets, J. S. Wurtele *and* B. A. Shadwick, **Phys. Plasmas 4**, 1872 (1997).

[21] P. Sprangle etal, **Phys. Rev. Lett. 69**, 2200 (1992).

[22] E. Esarey, J. Krall, and P. Sprangle, **Phys. Rev. Lett. 72**, 2887 ( 1994).

[23] C. D. Decker, W. B. Mori and T. Katsouleas**, Phys. Rev. E 50**, R3338 (1994).

[24] C. D. Decker etal, **Phys. Plasmas 3**, 1360 (1996).

[25] P. Sprangle etal , **Phys. Rev. Lett. 85**, 5110 ( 2000).

[26] N. E. Andreev and S. V. Kuznetsov, **Plasma Phys. Control. Fusion 45**, A39 (2003).

[27] C. I. Moore etal, **Phys. Rev. Lett. 79**, 3909 (1997).

[28] M. I. K. Santala etal, **Phys. Rev. Lett. 86**, 1227 (2001).

[29] S.-Y. Chen etal, **phys. Plasmas 6**, 4739 (1999).

[30] P. Bertrand etal, **Phys. Rev. E 49**, 5656 (1994).

[31] Z.-M. Sheng etal, **Phys. Rev. Lett. 88**, 055004-1 (2002).

[32] B. S. Paradkar etal, **Phys. Rev. E 83**, 046401 (2011).

[33] E. Khalilzadeh, A. Chakhmachi and J. Yazdanpanah, **Plasma Phys. Control. Fusion 59**, 125004 (2017).

[34] M. Pishdast, J. Yazdanpanah and A. Ghasemi, **Laser and Particle Beams 36**, 41 (2018).

[35] D. F. Gordon etal, **Phys. Rev. E 64**, 046404 (2001).

[36] K. Mima *etal*, **Phys. Plasmas 8**, 2349 (2001).

[37] C. J. McKinstrie and D. F. DuBois, **Phys. Fluids 31**, 278 (1988).

[38] C. Rousseaux etal, **Phys. Rev. Lett. 74**, 4655 (1995).

[39] S. Guérin *etal*, **Phys. Plasmas 2**, 2807 (1995).

[40] B. Quesnel etal, **Phys. Rev. Lett. 78**, 2132 (1997).

[41] H. C. Barr, S. J. Berwick, **Phys. Rev. Lett. 81**, 2910 (1998).

[42] H. C. Barr, P. Mason, and D. M. Parr, **Phys. Rev. Lett. 83**, 1606 (1999).

[43] J. Yazdanpanah, **Plasma Phys. Control. Fusion 60**, 025014 (2018).

[44] C. B. Schroeder etal, **Phys. Rev. Lett. 106**, 135002 (2011).

[45] P. Sprangle, E. Esarey, and A. Ting, **Phys. Rev. A 41**, 4463 (1990).

[46] S. V. Bulanov etal, **Phys. Fluids B 4**, 1935 (1992).

[47] C. D. Decker etal, **Phys. Plasmas 3**, 2047 (1996).

[48] G. B. Arfken and H. J. Weber, *Mathematical methods for physicists*, 6$^{th}$ Ed., (**Elsevier Academic Press**, New York, 2005).

[49] E. Esarey etal, **IEEE Transactions on Plasma Science 21**, 96 (1993).

[50] E. F. Toro, *Riemann Solvers and Numerical Methods for Fluid Dynamics: A Practical introduction*, 3$^{rd}$ Ed., (**Springer-Verlag**, Berlin, 2009).

[51] B. A. Shadwick, C. B. Schroeder and E. Esarey, **Phys. Plasmas 16**, 056704 (2009).

[52] X. Y. Hu, N. A. Adams, and C-W Shu, **J. Comput. Phys. 242**, 169 (2013).

[53] J. Yazdanpanah and A. Anvary, **Phys. Plasmas 19**, 03310 (2012).

[54] J. Yazdanpanah and A. Anvary, **Phys. Plasmas 21**, 023101 (2014).

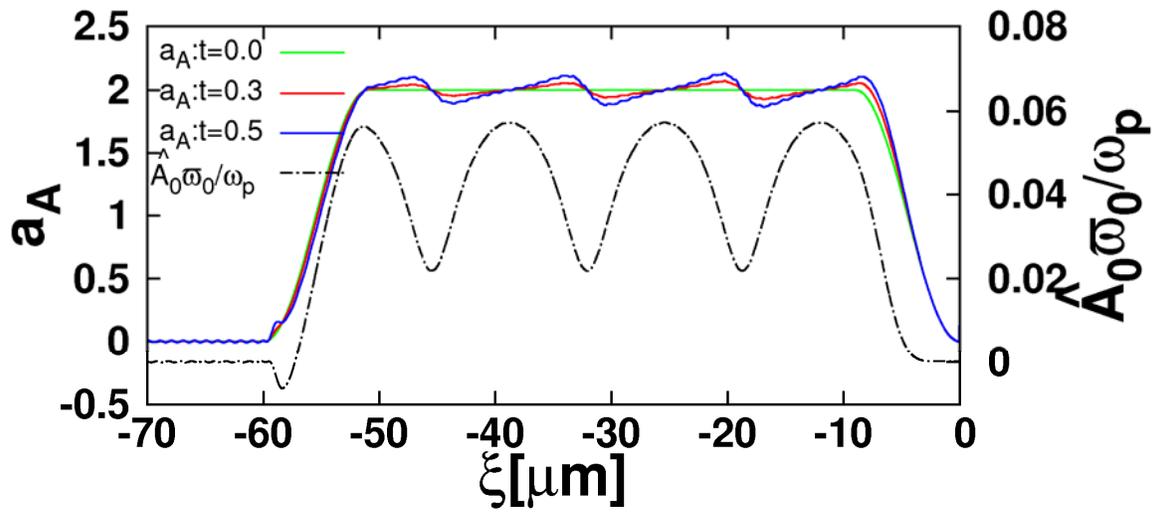

**FIG. 1**

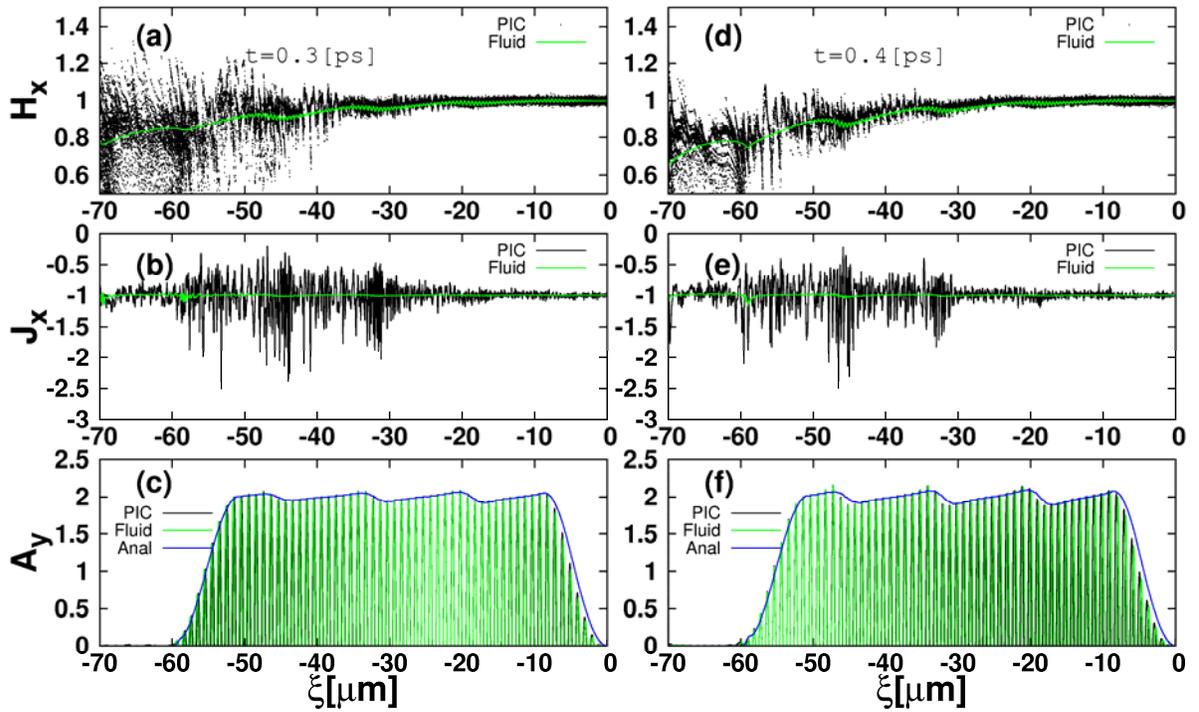

**FIG. 2**

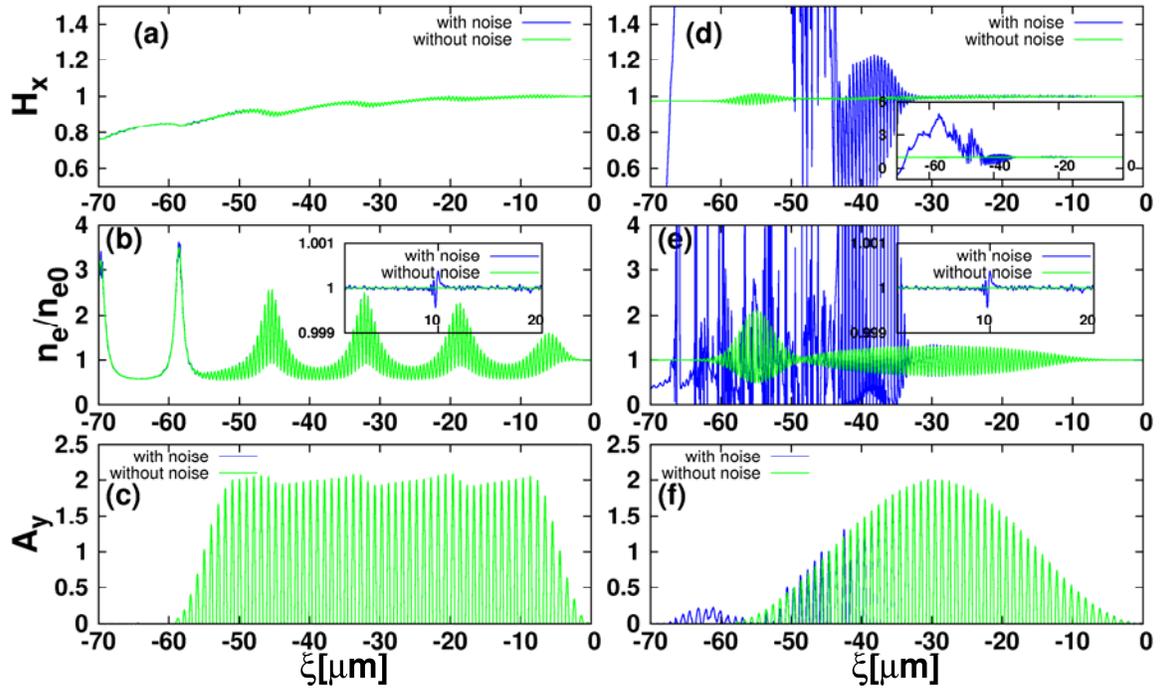

**FIG. 3**

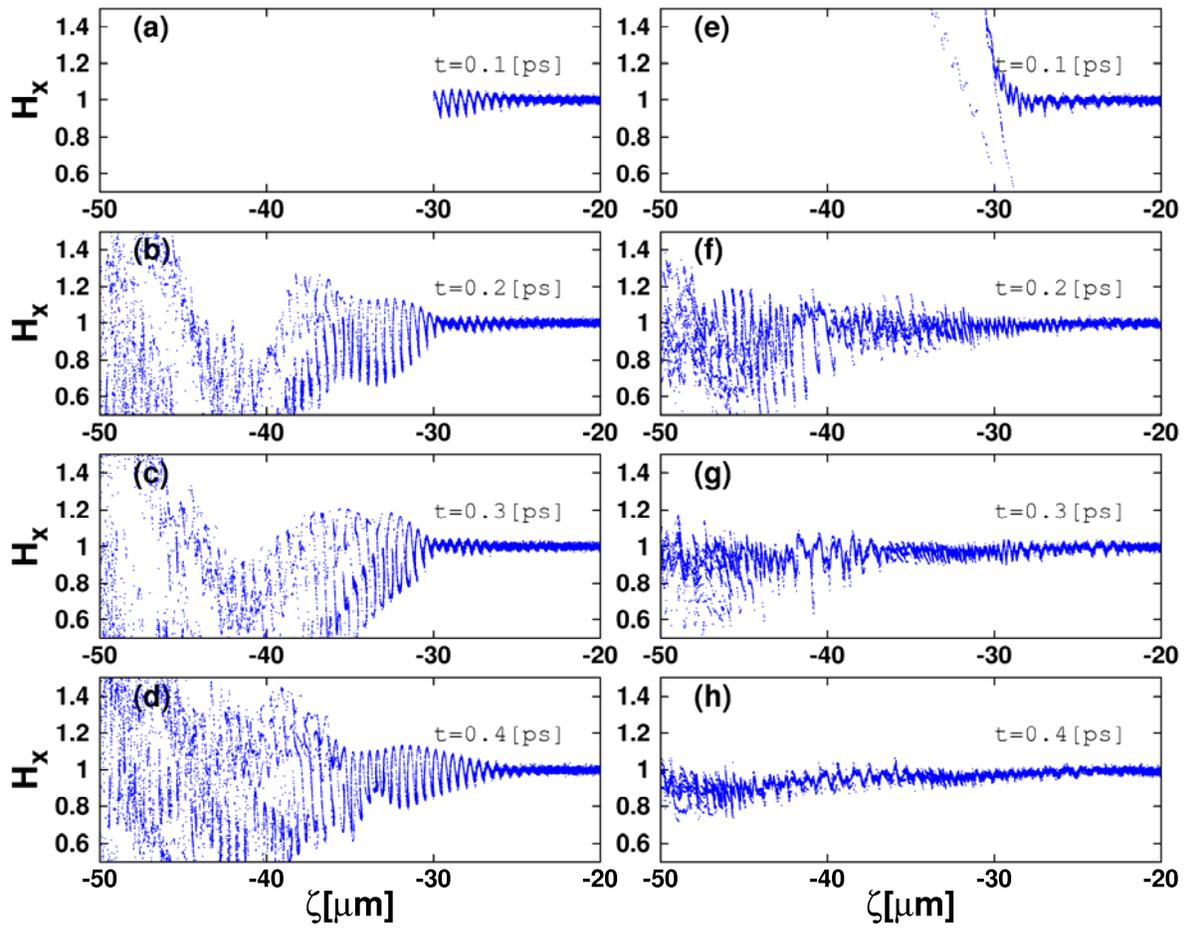

**FIG. 4**

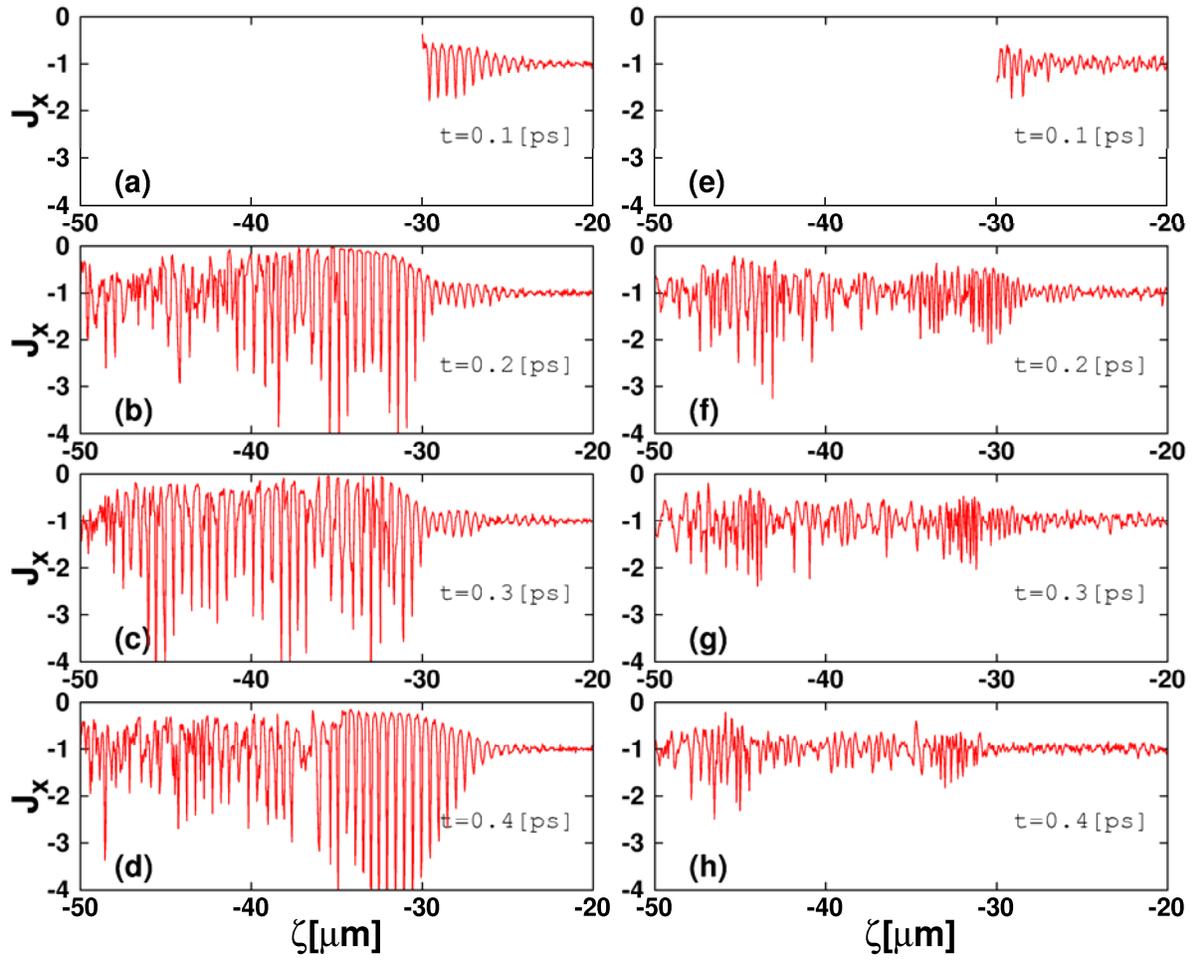

**FIG. 5**

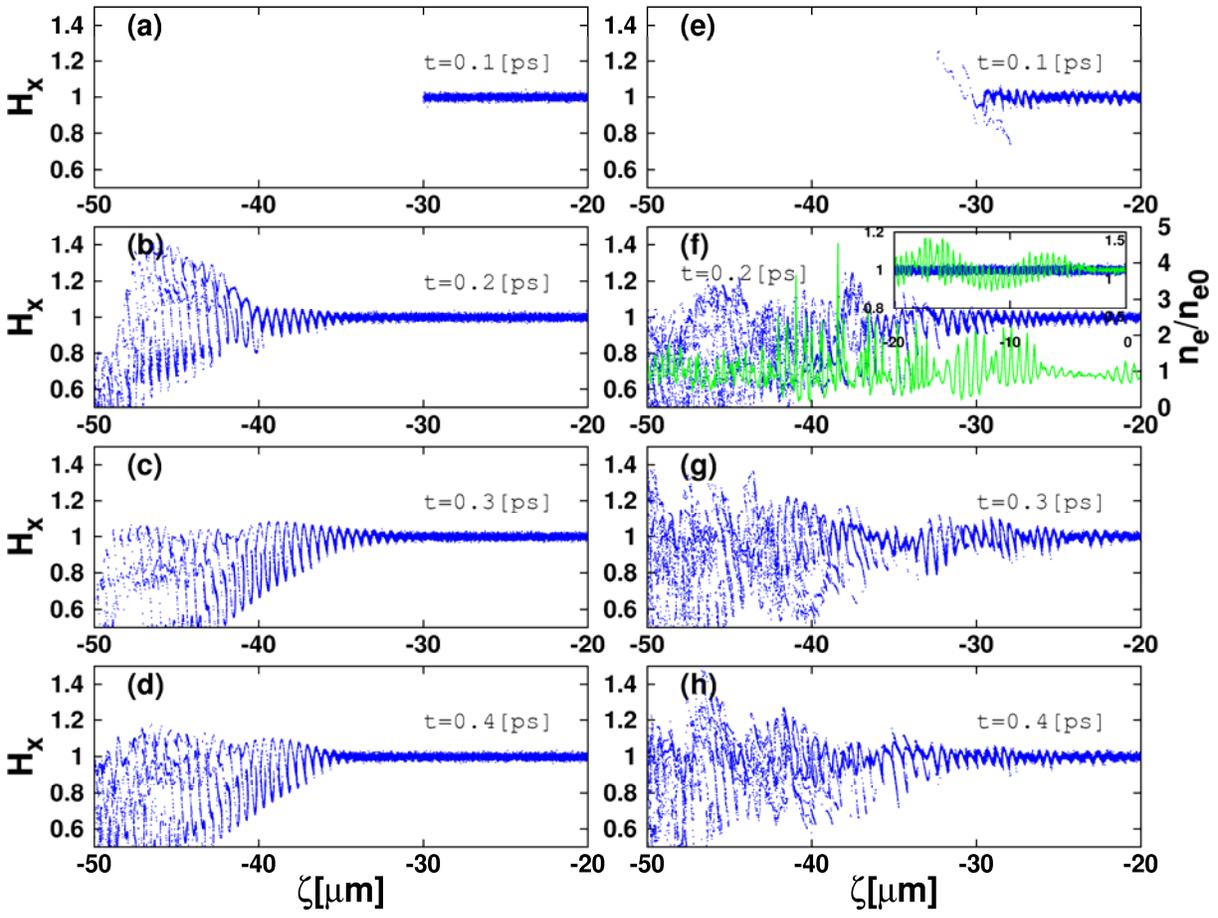

**FIG. 6**

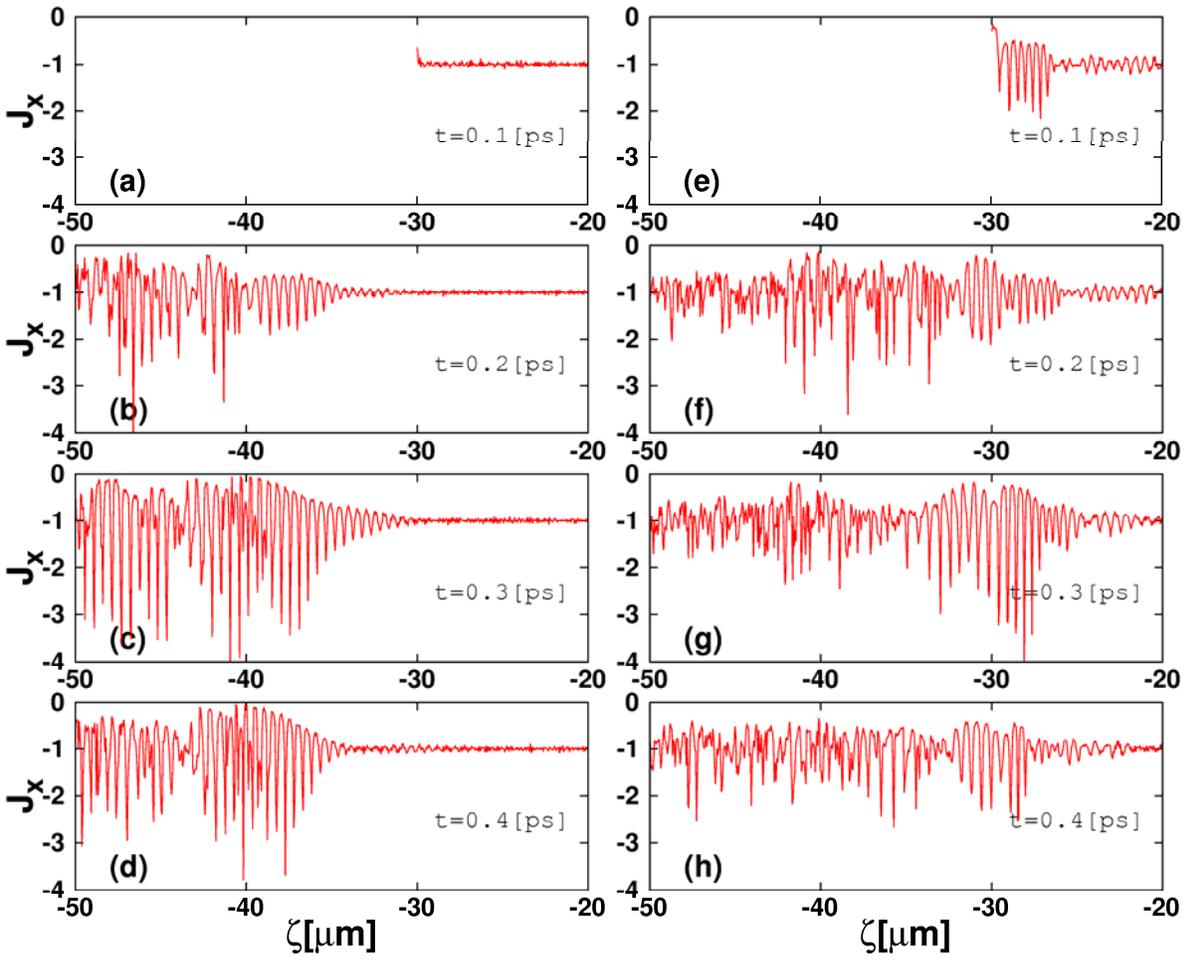

**FIG. 7**

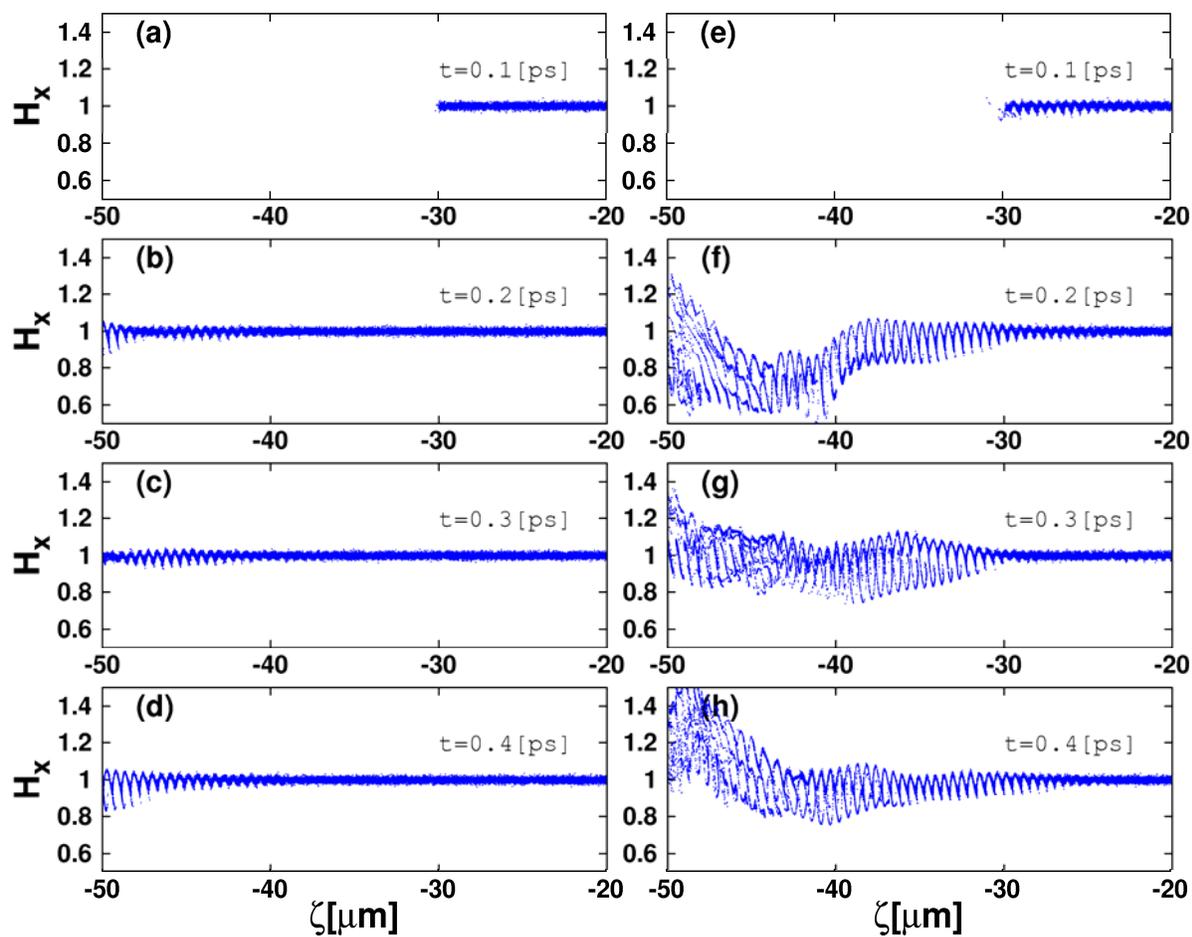

**FIG. 8**

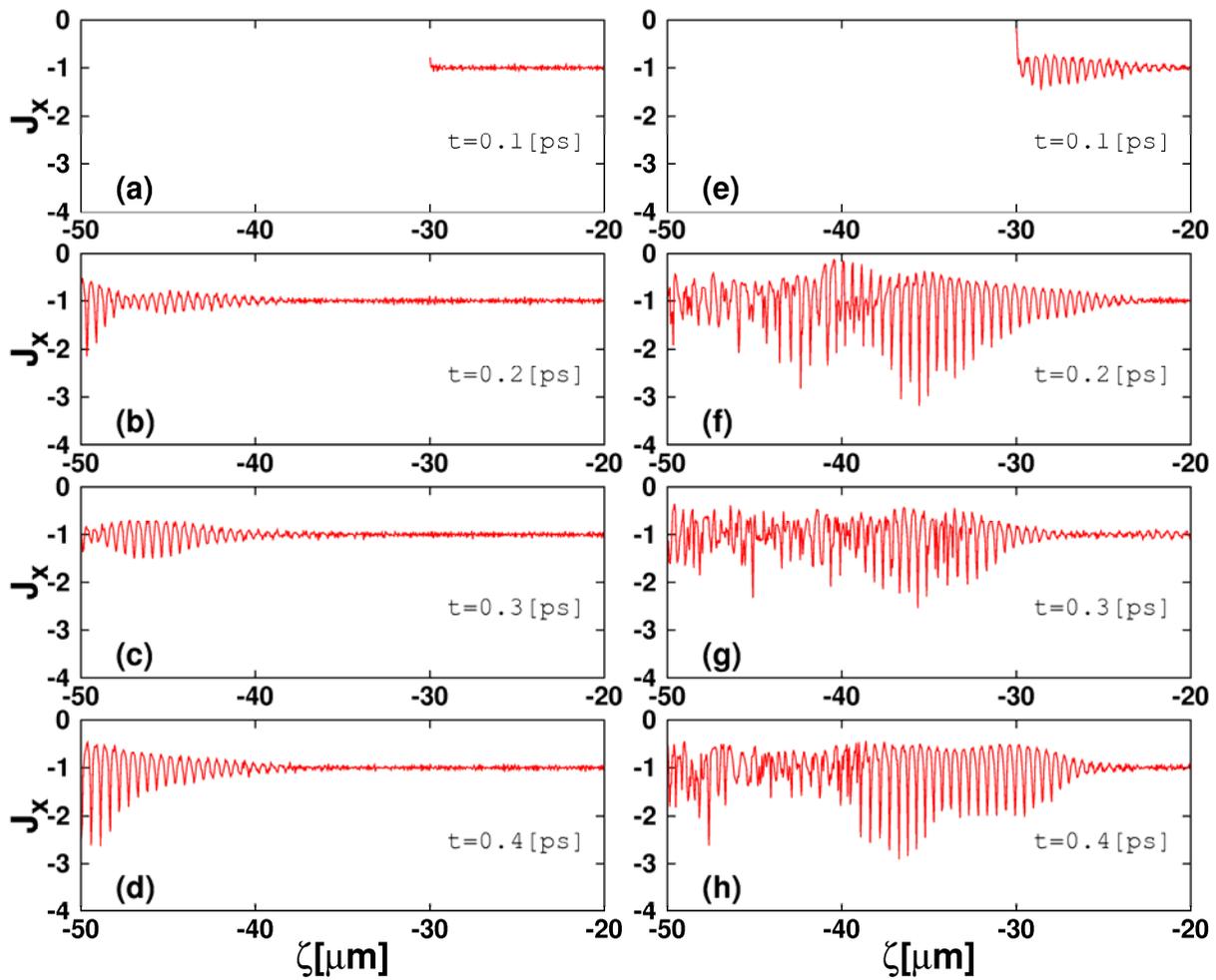

**FIG. 9**

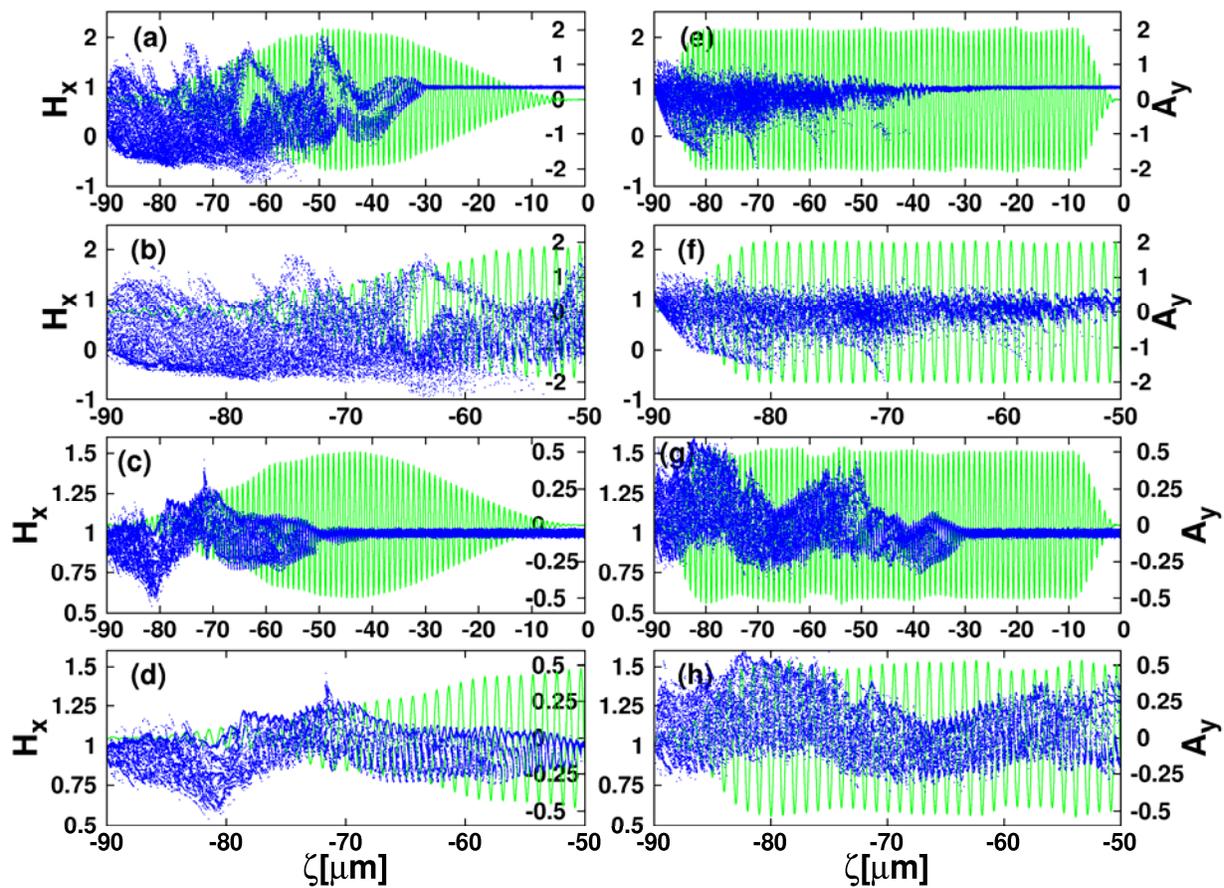

**FIG. 10**

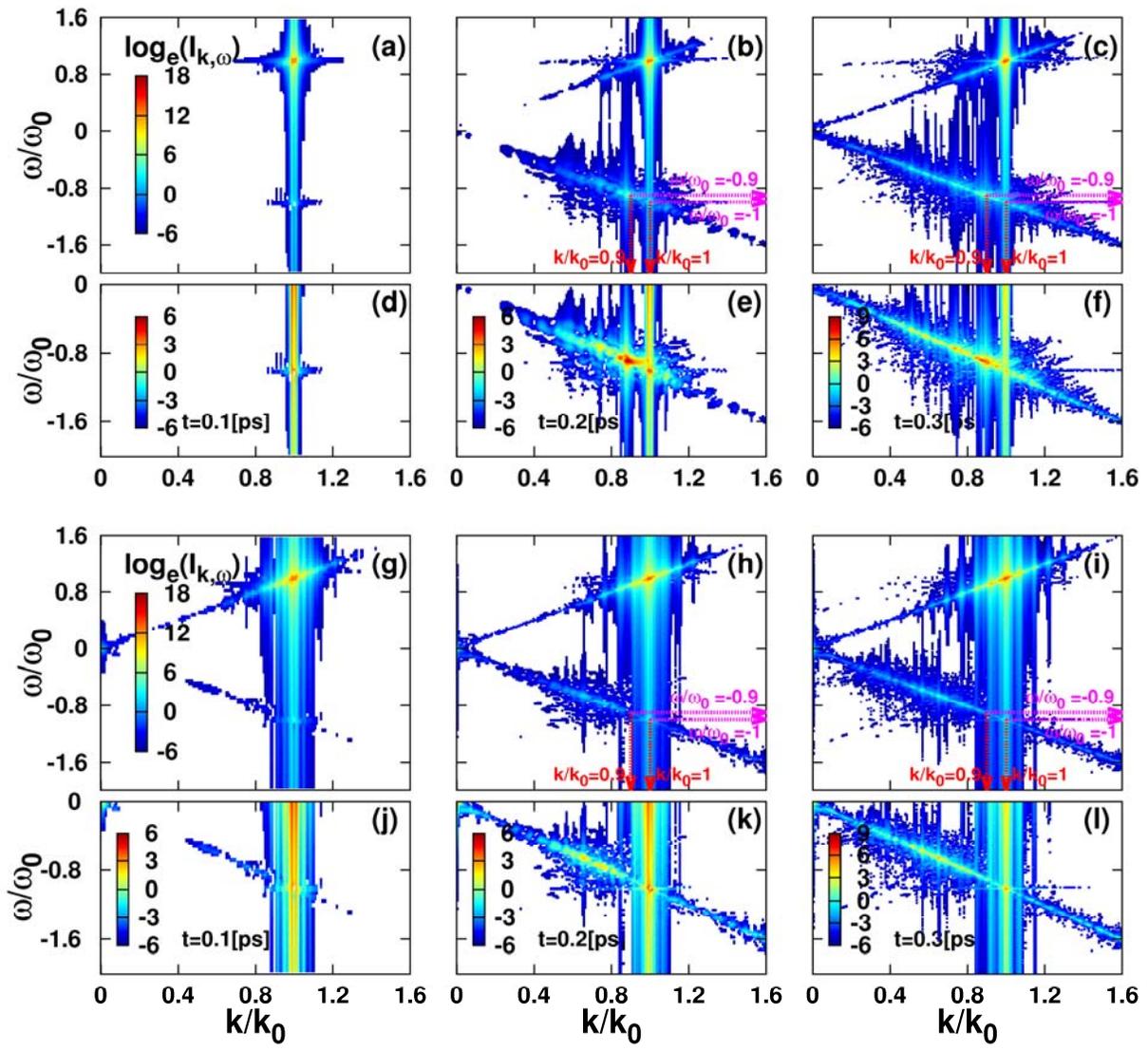

**FIG. 11**

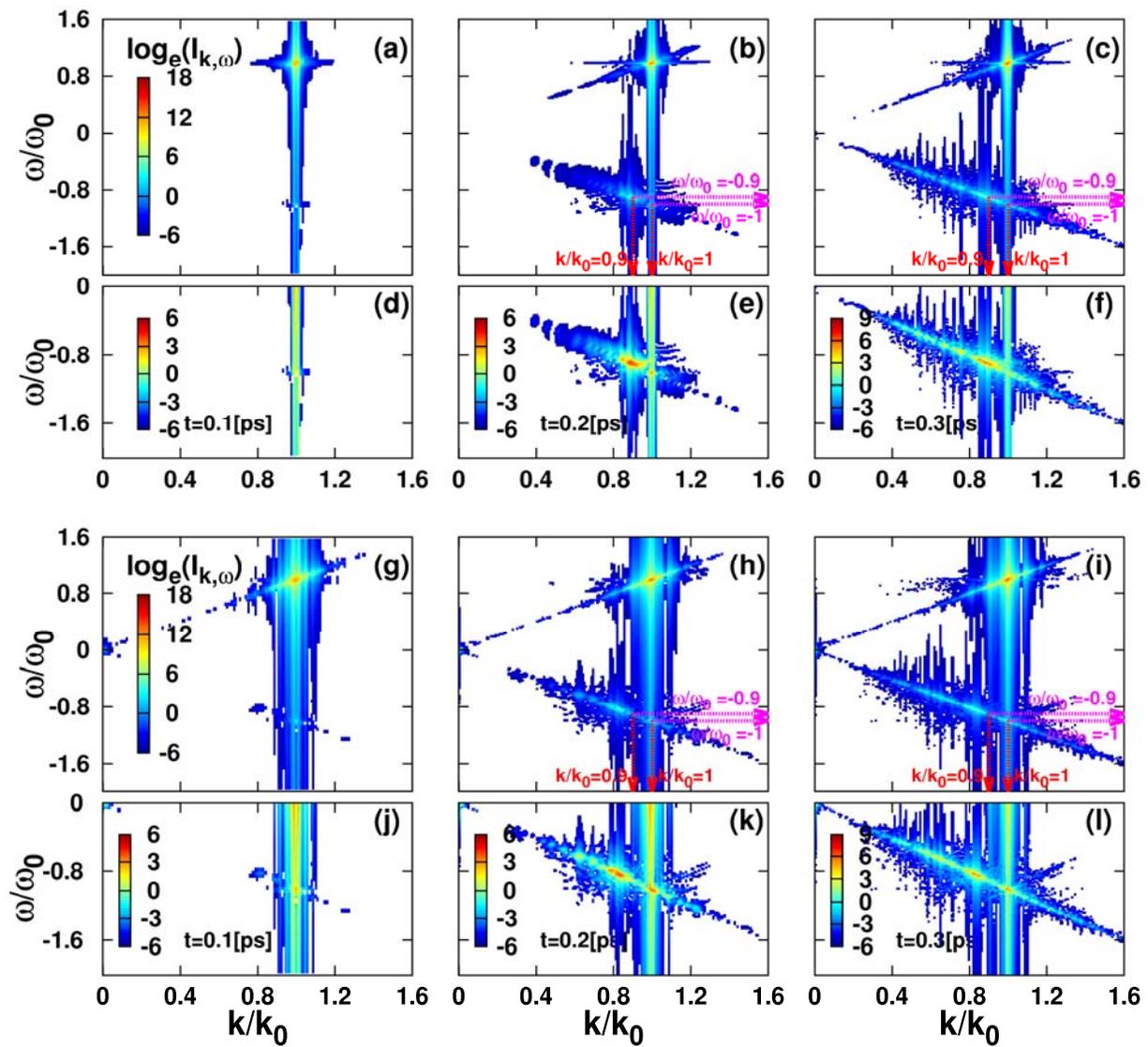

**FIG. 12**

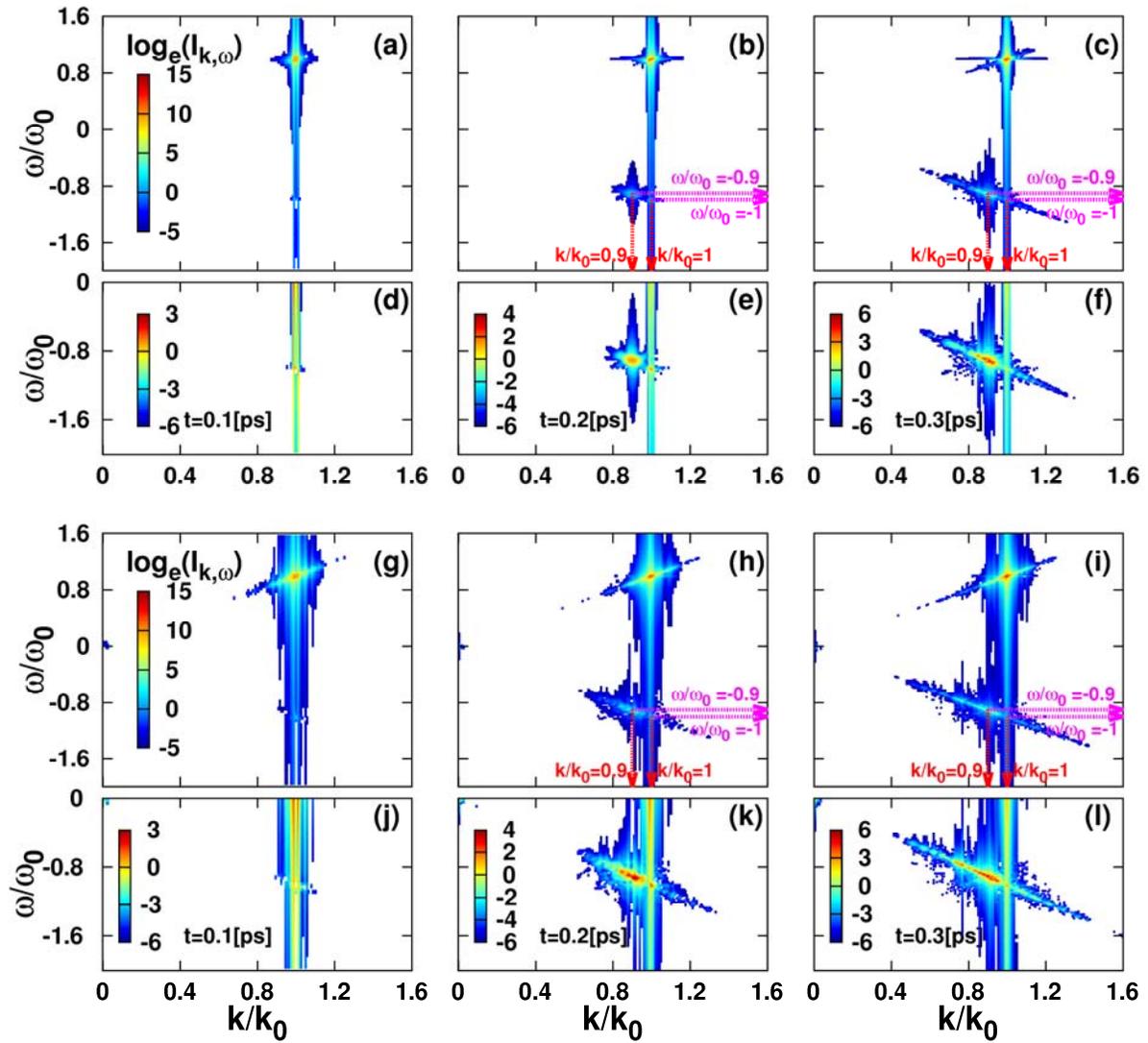

**FIG. 13**

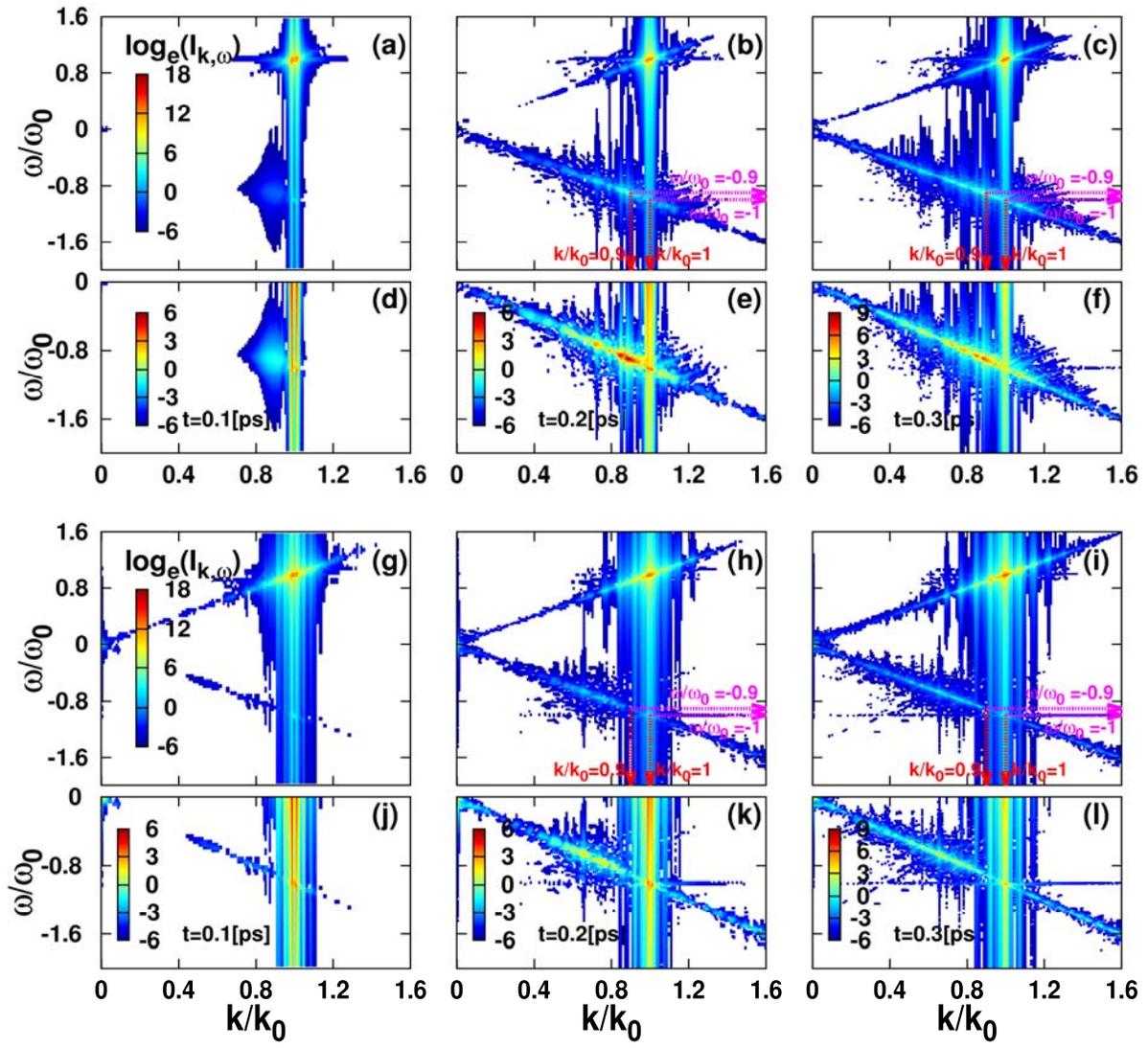

**FIG. 14**